\documentclass[11pt]{article}
\usepackage{geometry} 
\usepackage[parfill]{parskip} 
\usepackage{graphicx}
\usepackage{amsmath}
\usepackage{subfigure}
\usepackage{setspace}
\usepackage{epstopdf}
\usepackage{natbib}
\usepackage{times}

\begin{document}
  
  \title{\bf\Large Hydrodynamics of undulatory fish schooling in lateral configurations}
  \author{Li Jeany Zhang\thanks{lijeany@ucla.edu}~and Jeff D.~Eldredge\thanks{Corresponding author, eldredge@seas.ucla.edu} \\ Mechanical \& Aerospace Engineering Department \\ University of California, Los Angeles, 90095}
  
  
  \maketitle
  
  \begin{abstract}
    The thrust benefits of lateral configurations of two-dimensional undulating fish-like bodies are investigated using high-fidelity numerical simulation.  The solution of the Navier--Stokes equations is carried out with a viscous vortex particle method.  Configurations of tethered pairs of fish arranged side by side are studied by varying the lateral separation distance and relative phase difference.  It is shown that, in mirroring symmetry, the fish in the pair augment each other's thrust even at relatively large separations (up to ten body lengths).  At small distances, this augmentation is primarily brought about by a peristaltic pumping in the gap between the fish, whereas at larger distances, the thrust is affected by subtle changes in the vortex shedding at the tail due to interactions with the other fish.  In cases without symmetric undulation, one fish always draws more benefit from the interaction than the other.  Finally, lateral configurations with three fish are studied with mirroring symmetry between neighboring fish.  Whereas the center fish draws a net thrust benefit, this comes at the expense of a net drag on the outer two fish.  Each adjacent pair in this arrangement is slightly affected by the presence of the third fish.
  \end{abstract}
  
  \section{Introduction}
  
  The hydrodynamics of fish locomotion at moderate Reynolds number has been of interest to biologists and engineers for decades.  It is known that unsteady rhythmic undulations of fish `control surfaces' -- the body, fins and tail -- play a fundamental role in enabling a fish to propel itself through the water \citep{fisharfm:1j}.  This propulsion is achieved primarily by the reaction force supplied by the fluid against the accelerating and deforming surfaces.  Theoretical models for fish swimming have been proposed by \citet{lighthillelong:2j, lighthillelong:3j} and \citet{WuT:1j}, which utilize slender body theory to explain thrust generation in fish-like swimming.  More recently, it has been found that the mechanism of vorticity shedding off the fins and tail is an important feature in fish locomotion \citep{triantaarfm00:1j, triantafyllou:1j}.  The near-field flow structure, wake structure and body vortices around the fish skin have been studied both experimentally and computationally \citep{LiuH:1j, ZhuQ:1j, Clark:1j}.  Moreover, Triantafyllou and co-workers \citep{triantafyllou:1j, triantafyllou:2j,triantaarfm00:1j} have investigated the range of Strouhal number in flapping foils and fish-like systems that lead to peak propulsive efficiency.  Propulsive efficiency is defined as $\eta = \overline{T}\,\overline{U}/\overline{E}$ in Triantafyllou's paper, where $\overline{T}$ is the time-averaged thrust, $\overline{U}$ is the average forward velocity, and $\overline{E}$ is the average input power.  In their work, the range $0.2\,<\,St\,<\,0.4$ usually guarantees the highest propulsive efficiencies.  This Strouhal number range is also seen in aquatic creatures, such as sharks and cetaceans \citep{RohrJJ:1j, triantafyllou:1j, triantafyllou:2j, triantaarfm00:1j}. 
  
  However, these studies are all based on individual fish.  Indeed, many kinds of fishes tend to swim in groups \citep{WeihsD:1j, HunterJR:1j, VanOlst:1j}.  A number of explanations for fish schooling have been advanced, for example, protection against predation \citep{Cushing:1j}, social animal behaviors \citep{PitcherTJ:4j}, and reduction of navigation errors \citep{LarkinPA:1j}.  In particular, it is thought that energy savings through hydrodynamic interactions could be an important factor.  \citet{WeihsD:1j} postulated that the hydrodynamics of fish schooling is strongly tied with both streamwise and lateral interactions among the fish, and proposed an optimal configuration in which members would draw the most energetic benefit.   Though there has been some skepticism about the relative importance of achieving hydrodynamic benefit in schooling \citep{Partridge:3j}, it is reasonable to expect that some benefit is achieved, even after the necessary compromises that detract from energy savings to fulfill the other roles of the school \citep{AbrahamsMV:1j}.  \citet{Herskin:1j} have found evidence of energy savings by comparing oxygen consumption rates and tail-beat frequencies between members at the front and rear of the school.  More recently, \citet{Noren:1j} examined cetacean mother--calf pairs swimming in echelon formation and found that the calf was able to reduce its effort and increase its speed compared to solo swimming.  
  
  In contrast to the extensive investigations on self-propulsion mechanisms for individual fish-like swimming, there have been relatively few studies -- either computational or experimental -- on the hydromechanical consequences of fish schooling.  Studies on streamwise interactions have recently been performed in the context of fish swimming in an obstacle wake \citep{Liao:1j, Liao:2j, Beal:1j,eldredgeJFM:3j}.  \citet{eldredgeJFM:3j} computationally studied the passive self-propulsion of a three-link fish-like swimmer by focusing on the effect of the body length and flexibility of the fish.  However, lateral interactions among fish-like bodies with undulatory waves propagating from the head to the tail have not been explored thoroughly.  A notable exception is the recent computational study of \citet{DongGJ:1j}, in which an infinite lateral series of two-dimensional undulating fish-like shapes was investigated.  Each member of the series moved either in phase or exactly out of phase with its immediate neighbors.  They found that the anti-phase cases led to maximum thrust generation, while in-phase cases exhibited the greatest energy savings.  They found that benefits were insignificant for separation distances beyond one body length.
    
  Similar to the study of \citet{DongGJ:1j}, the present paper focuses on the hydrodynamics of two-dimensional fish-like shapes in lateral arrangement, undergoing undulatory motions in a viscous and incompressible flow.  However, in contrast to their work, the present investigation focuses on the thrust consequences in configurations with a finite number of individuals, over a wider range of separation distances and a variety of phase differences.  The study relies on numerical simulations of the Navier--Stokes equations with the viscous vortex particle method \citep{jdevvpm:1j, ZhangEldredge:1j}.  Single and multiple (up to three) fish tethered in a free-stream flow are simulated in this work.  The Reynolds number  -- the ratio of inertial to viscous forces -- is maintained relatively low at 100 for computational benefits.  However, the results can be extended to higher Reynolds number, since the reactive mechanisms at the heart of the interaction are insensitive to this parameter.  The Strouhal number, which represents the ratio of tail undulation velocity to free-stream velocity, is varied in a single fish to determine the value at which thrust balances drag.  This Strouhal number is then fixed in all lateral configurations of fish-like bodies.  The hydrodynamics of lateral schooling of fish-like bodies is studied by varying the separation distance and phase difference between adjacent fishes. Finally, a lateral configuration with three fishes is also investigated here.    

It is noted that, although this study is only focused on an abstracted model for fish-like swimming, the purpose of this study -- to examine hydrodynamic interactions between nearby fish-like shapes and their effect on thrust -- is also relevant for understanding similar mechanisms in real fish.  Furthermore, this study represents the first step in a logical sequence towards a more comprehensive understanding of hydrodynamic interactions.
  
  \section{Materials and methods}
  
  The present investigation makes use of the viscous vortex particle method \citep{Koumoutsakos:1j,Ploumhans:2j,jdevvpm:1j,cottetbook:1j} to simulate the flow.  This method solves the Lagrangian form of the Navier-Stokes equations by utilizing vorticity-carrying particles as computational elements.  The no-slip boundary condition is enforced by solving for and fluxing a surface vortex sheet representing spurious slip into adjacent particles in the fluid.  The method was recently extended to continuously deforming bodies by \citet{ZhangEldredge:1j}.  Convergence of the overall algorithm to the Navier-Stokes solution was demonstrated in their work.  The reliance on convecting computational particles rather than a grid affords this method with a natural adaptivity to the evolving shape of the fish.  Furthermore, there is no difficulty in simulating multiple fishes, even if separated by tens of body lengths.  Particles interact with each other through the Biot-Savart summation, which is accelerated by a fast multipole method \citep{Carrier:1j}.  Details of the VVPM including force calculation for deforming bodies, including convergence studies, can be found in the paper by \citet{ZhangEldredge:1j}.

\begin{figure}[p]
    \centering
    \includegraphics[scale=0.45]{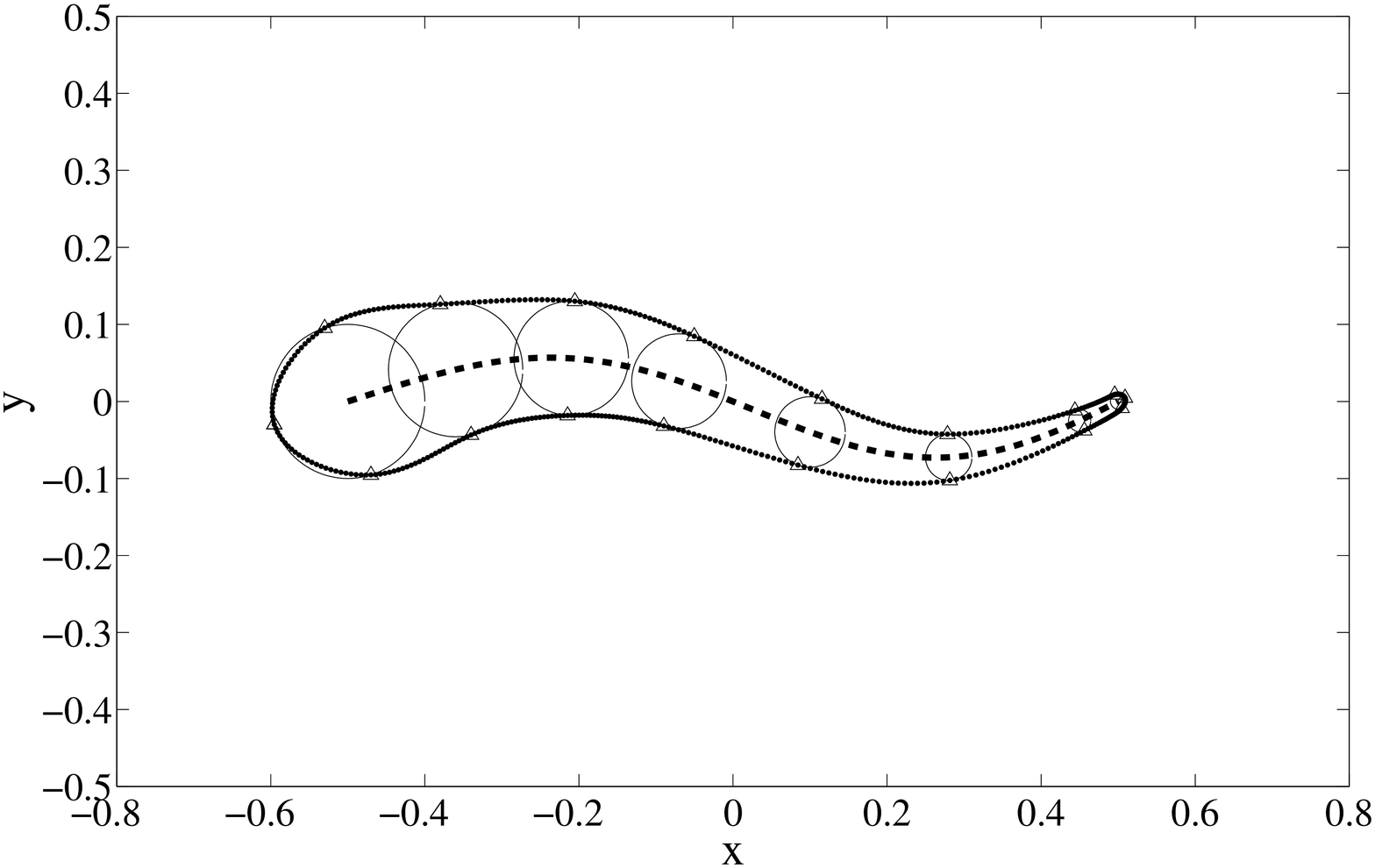}
    \caption{Schematic of the fish design: $--$ backbone, $\triangle$ attachment points, $-$ skin.}
    \label{fig:FishDesign}
  \end{figure}

The profile of each is prescribed at all times.  Its construction, depicted schematically in Fig.~\ref{fig:FishDesign}, is based on the time-varying shape of a backbone curve. Several circles of different radius are centered on the backbone, and cubic splines are drawn between points of tangency on these circles to represent the fish profile.  Further details can be found in the paper by \citet{ZhangEldredge:1j}.  

  The backbone undulatory motion is given by
  \begin{align}
    x_{c}(s,t) &= -0.5 + s, \\
    y_{c}(s,t) &= A_{0}e^{ks}\sin[2\pi(s-ft)],
  \end{align}
  where the parameter $s$ varies from 0 to 1, $A_{0} = 0.05$ is the undulation amplitude, $k = 0.5$ is a growth factor, $f$ is the backbone frequency and the period is $T = 1/f$.  This form, which is motivated by the analysis of carangiform fish mechanics by \citet{lighthillelong:1j}, produces a traveling wave of growing amplitude that propagates from the head to the tail.  By using this construction of a fish-like shape, the fractional change of area is never greater than 0.2 percent, thus the body configuration can be regarded as area preserving.
  
  It is known that tail-beat frequency is a key parameter in the propulsion mechanism of fish \citep{andertrianta:1j, fisharfm:1j}.  This effect is characterized in this work by the Strouhal number, defined as $St = fA_{T}/U_{\infty}$, where $f$ is the backbone frequency, $A_{T} = 2A_{0}e^{k}$ is the peak-to-peak tail amplitude, and $U_{\infty}$ is a free-stream velocity.  The Reynolds number, defined as $U_{\infty}L/\nu$, where $L$ is the horizontal chord length between the head and the tail (approximately 1.1 in our fish) and $\nu$ is the kinematic viscosity, is held fixed at $Re = 100$.  Finally, it is conventional to represent the force by a dimensionless force coefficient  The $x$ component of this force coefficient is defined as $C_{x} = F_{x}/(\frac{1}{2}\rho U_{\infty}^{2} L)$.  

  \section {Results}
  
  This section presents the computational results of a single two-dimensional fish-like shape as well as two and three fish-like profiles in lateral arrangement.  For simplicity, we denote this single fish-like shape as a single `fish' and the system with two and three fish-like bodies as `fish schooling'.  The first problem consists of performing a study to understand the effect of Strouhal number on thrust generation in a fixed solo fish which is undergoing prescribed undulatory deformation in a uniform free stream.  The second problem comprises an investigation of the thrust improvement mechanism in the lateral configuration of two and three fish schooling with respect to changes in separation distance and phase.
  
  \subsection{Tethered single fish in an uniform free stream}
  
\begin{figure}[t]
    \centering
    \includegraphics[scale=0.45]{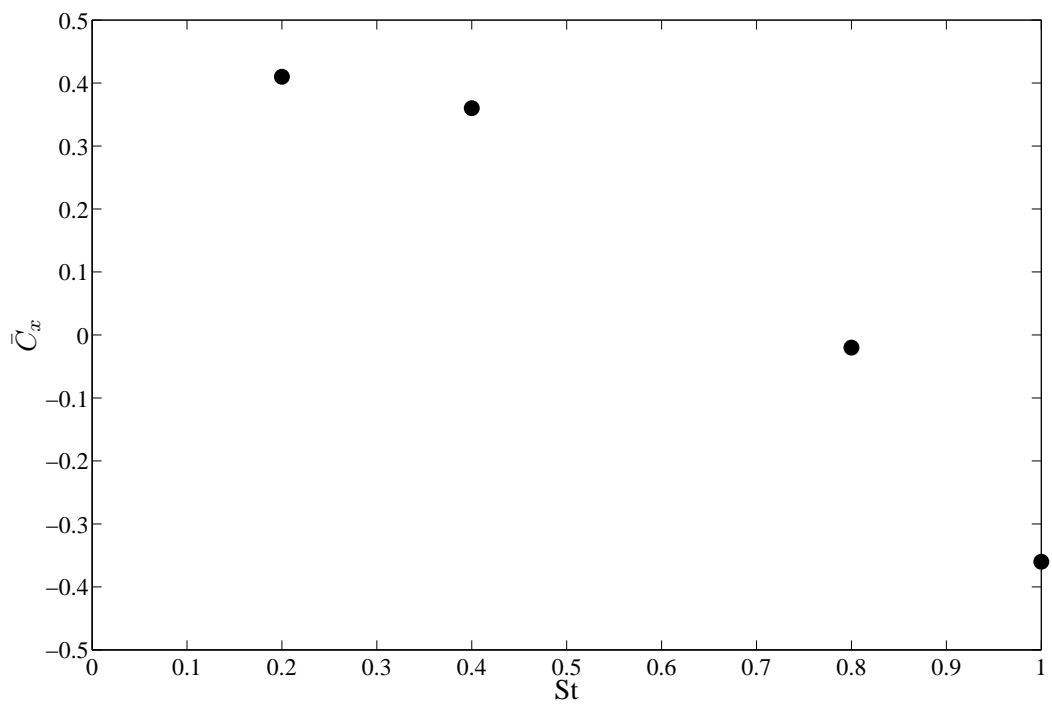}
    \caption{$x$ component of mean force coefficient versus Strouhal number in the case of solo fish.}
    \label{fig:SoloFish_Cx_St}
  \end{figure}

\begin{figure}[t]
    \centering
    \includegraphics[scale=0.5]{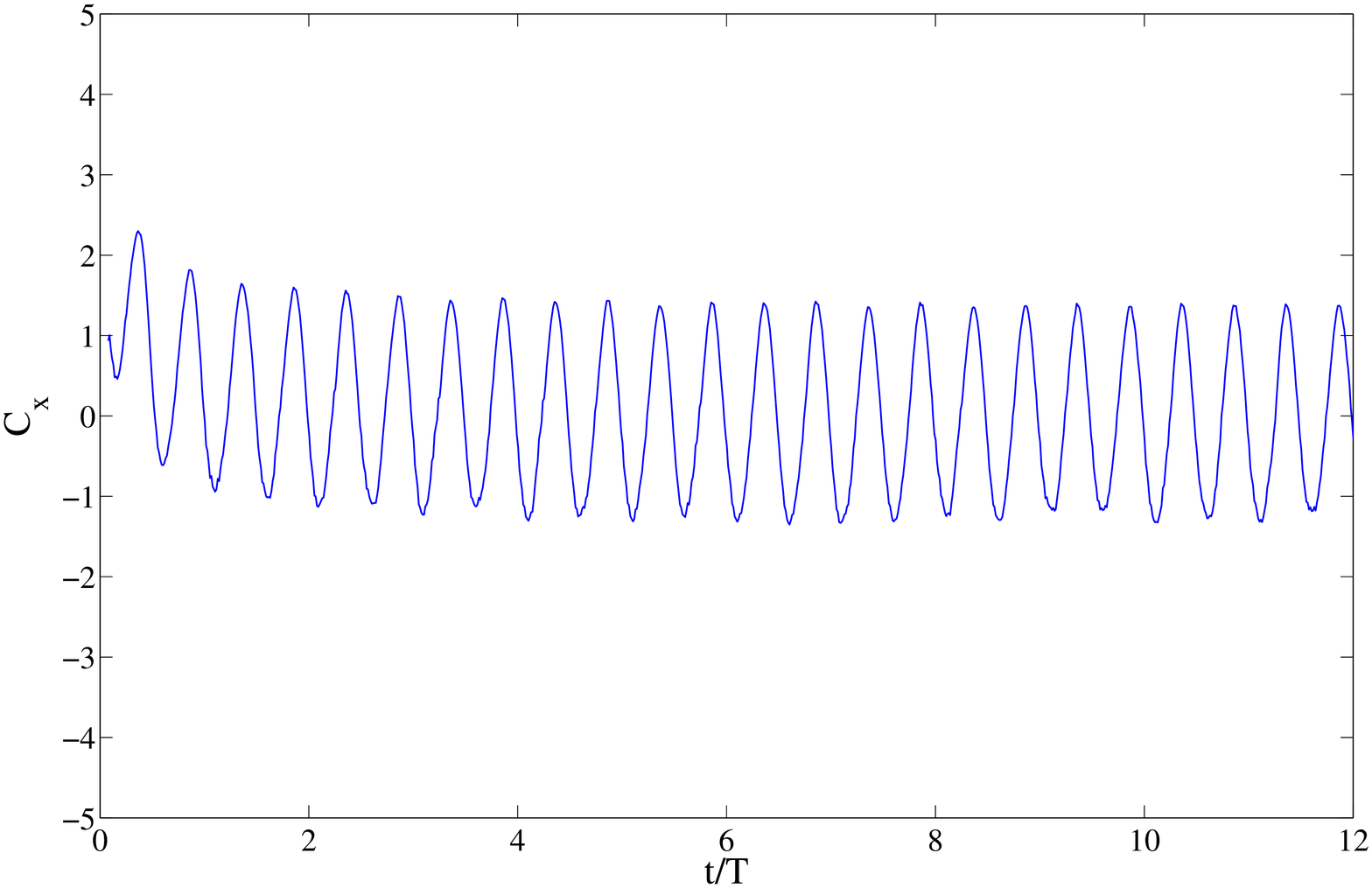}
    \caption{Time-varying $x$ component of force coefficient in the case of solo fish with $St = 0.8$.}
    \label{fig:SoloFish_Cx_T}
  \end{figure}

  In order to understand mechanisms for thrust augmentation in fish schooling, we firstly explore the two-dimensional flow and the resulting force produced by an undulating fish-like profile immersed in a uniform free stream.  The Strouhal number is varied over a range of values between 0.2 and 1.  Fig.~\ref{fig:SoloFish_Cx_St} reveals a smoothly declining trend of the mean $x$ component of the force coefficient $\overline{C}_{x}$ as Strouhal number increases.  In particular, it is noted that $St = 0.8$ gives nearly zero mean net force and above this Strouhal number, the fish generates a net thrust.  Thus, $St = 0.8$ indicates the condition at which thrust balances drag in the mean sense, and the equivalent free-swimming fish would generate nearly the same flow.  An example of the time-varying history of force coefficient is shown in Fig.~\ref{fig:SoloFish_Cx_T} for $St = 0.8$.  Consequently, in the following cases of lateral arrangements of two and three fish schooling, Strouhal number remains fixed at $St = 0.8$ to demonstrate thrust improvement mechanisms induced by schooling configurations. 

  It is useful to compare this Strouhal number with those found in previous studies of aquatic creatures.  In our study, Strouhal numbers greater than 0.8 are needed to produce net thrust, which is higher than those found in the studies of Triantafyllou and co-workers \citep{triantafyllou:1j, triantafyllou:2j, triantaarfm00:1j}.  This can be mainly attributed to the low Reynolds number ($100$) used in our study, while Trianatafyllou's papers are focused on high Reynolds numbers of and above order $10^{4}$.  In order to achieve the same thrust, low Reynolds number needs to be coupled with high Strouhal number due to the increased role of viscous drag.  Most studies in fish locomotion are based on high Reynolds number which are typical of aquatic animals, such as dolphins, but low Reynolds number is still widely seen in swimming creatures with smaller size, for example, tadpole larvae swimming is at Reynolds number on the order of $10^{2}$ \citep{LiuH:1j}.  In addition, spanwise vorticity shedding and other three-dimensional effects, which are missed here, also affect the thrust generation mechanism.  

  \subsection{Lateral schooling of two fish}

\begin{figure}[t]
    \centering
    \includegraphics[scale=0.45]{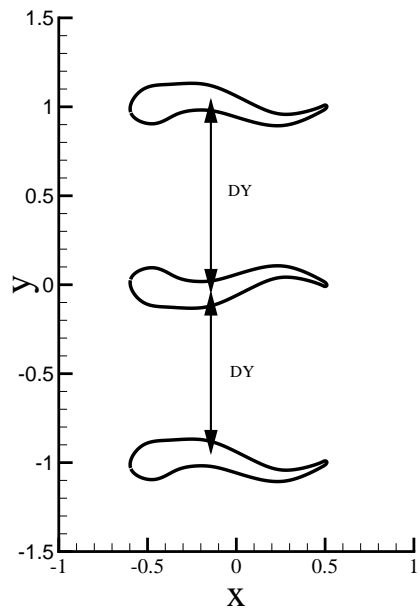}
    \caption{Schematic of lateral fish schooling with phase difference $\phi = \pi$ between adjacent fish.}
    \label{fig:FishSketchMap}
  \end{figure}

  In this section, we apply the same two-dimensional fish-like profile utilized in the previous solo fish study in lateral arrangements of two fishes, and simulate two classes of cases which both have fixed Strouhal number $St = 0.8$ and Reynolds number $Re = 100$.  The first class has fixed phase difference, which is set to be $\phi = \pi$, but different separation distance $\Delta Y$ defined as the distance between geometrical centroids on the upper and lower fish, $\Delta Y \in \{1, 1.5, 3, 4, 5, 7, 8, 10, 15\}$.  The other class has fixed separation distance $\Delta Y = 1.5$, but different phase difference, $\phi \in \{0.25\pi, 0.5\pi, 0.75\pi, \pi\}$.  Both cases are compared with the solo fish, which has the same periodically undulating deformation immersed in the unit free stream.  The schematic of lateral configuration is depicted in Fig.~\ref{fig:FishSketchMap} for three fishes; the two-fish configuration is similar. 
  
  \subsubsection{Effects of separation distance $\Delta Y$}
  
\begin{figure}[t]
    \centering
    \subfigure[$t/T = 0$]{
      \includegraphics[scale=0.2]{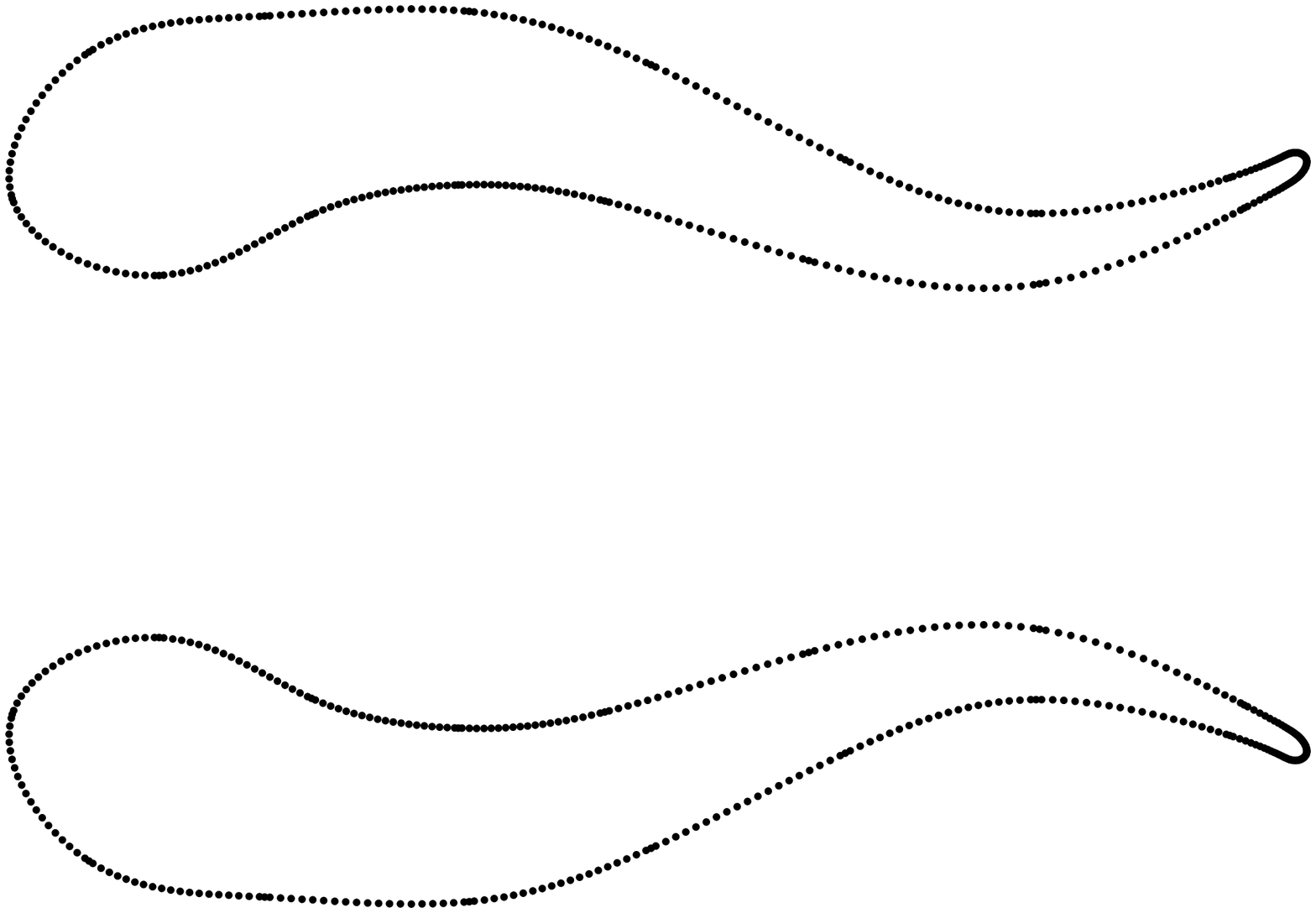}
    }
    \subfigure[$t/T = 1/4$]{
      \includegraphics[scale=0.2]{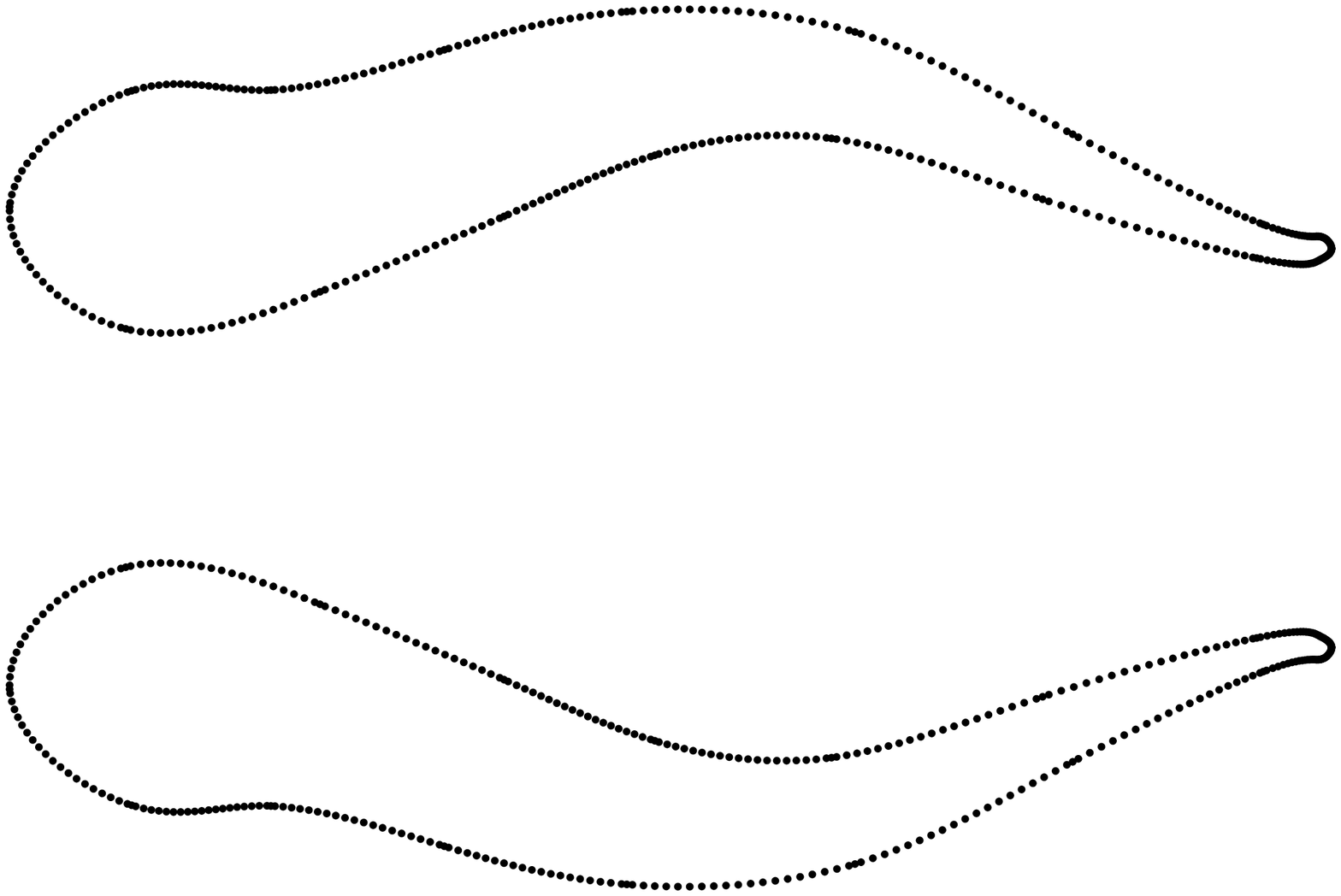}
    }
    \\
    \subfigure[$t/T = 1/2$]{
      \includegraphics[scale=0.2]{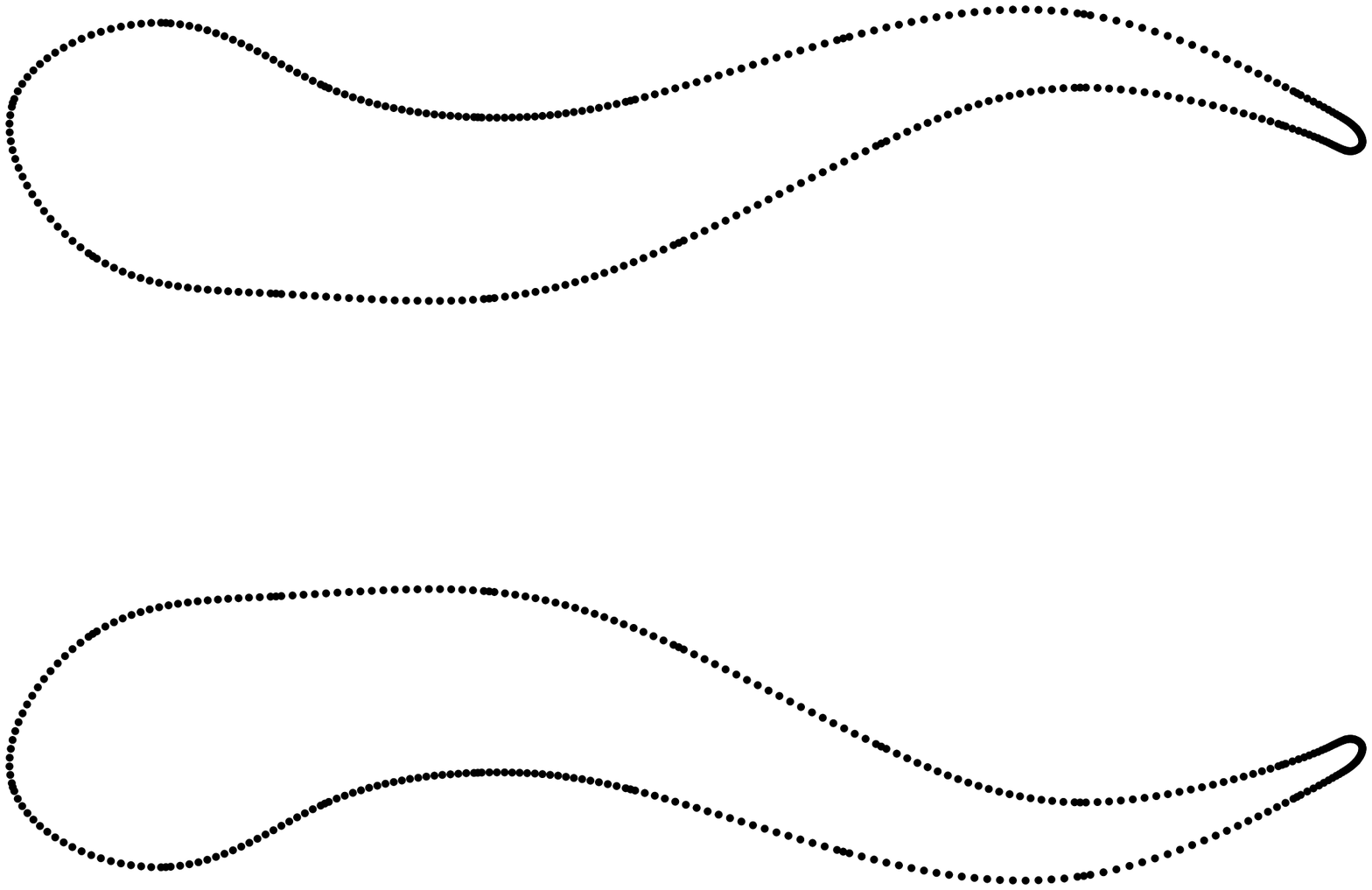}
    }
    \subfigure[$t/T = 3/4$]{
      \includegraphics[scale=0.2]{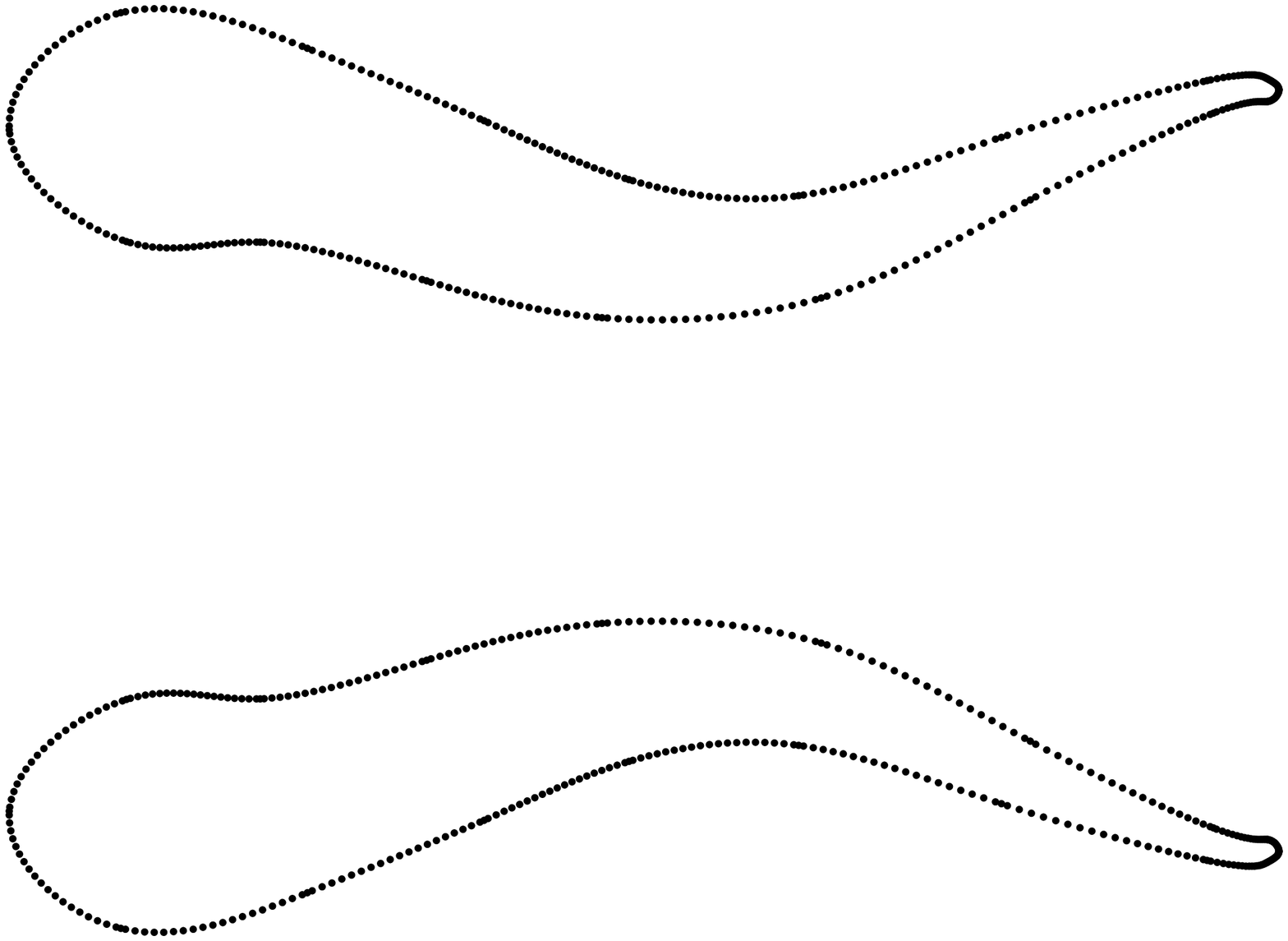}
    }
    \\
    \subfigure[$t/T = 1$]{
      \includegraphics[scale=0.2]{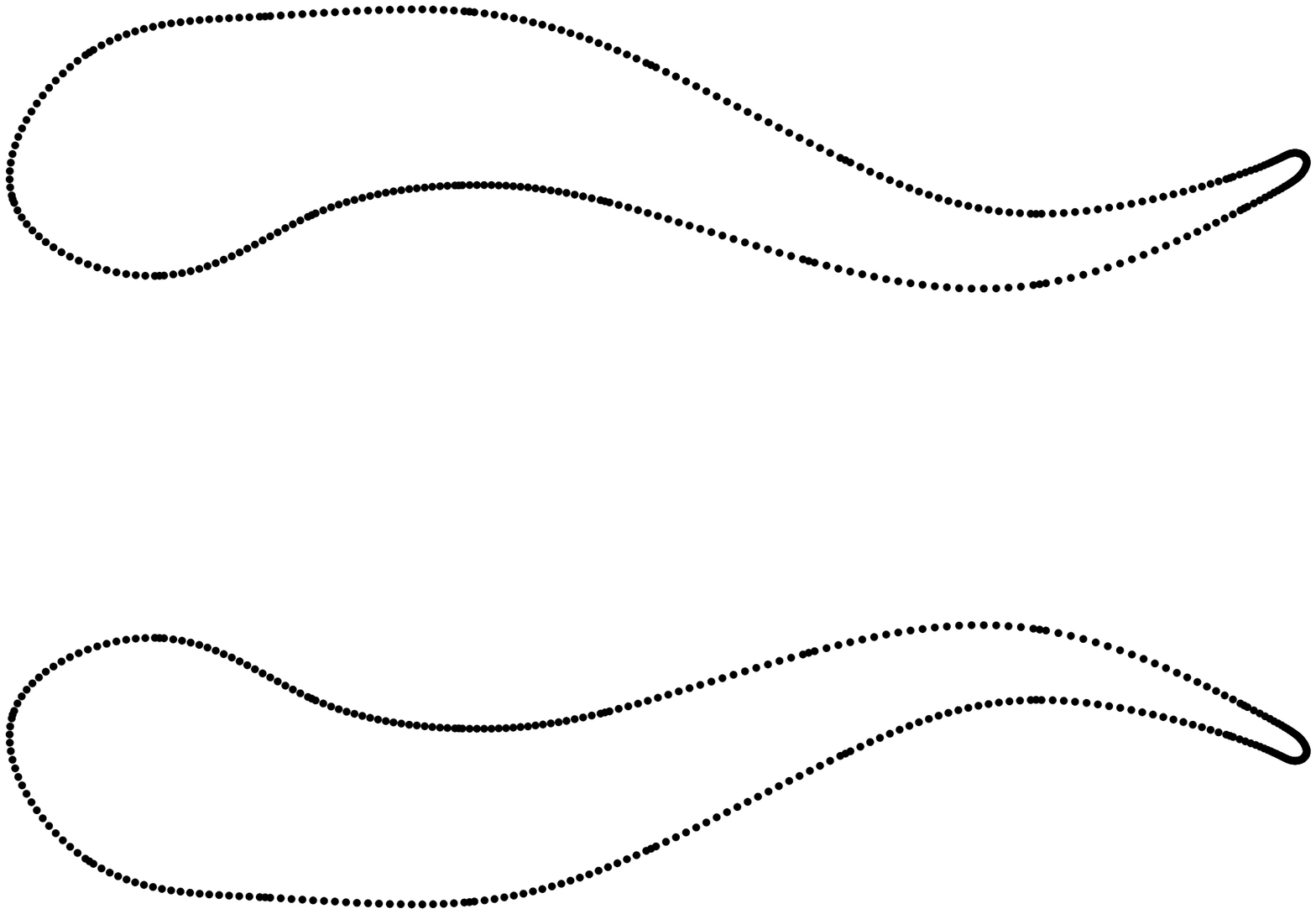}
    }
    \caption{Schematic of shape variation of lateral double fish schooling in one period with phase difference $\phi = \pi$.}
    \label{fig:FishShapeChange}
  \end{figure}

\begin{figure}[t]
    \centering
    \subfigure[Total force coefficient]{
      \includegraphics[scale=0.25]{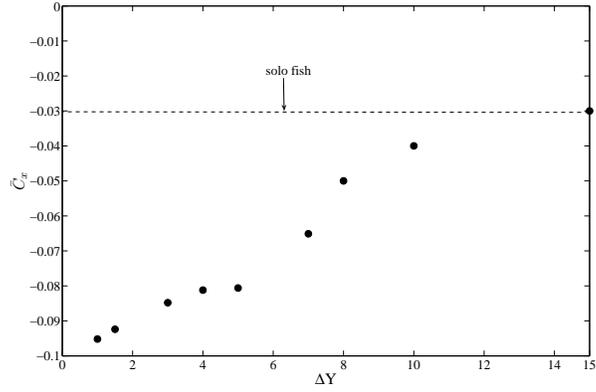}
    }
    \\
    \subfigure[Pressure force coefficient]{
      \includegraphics[scale=0.25]{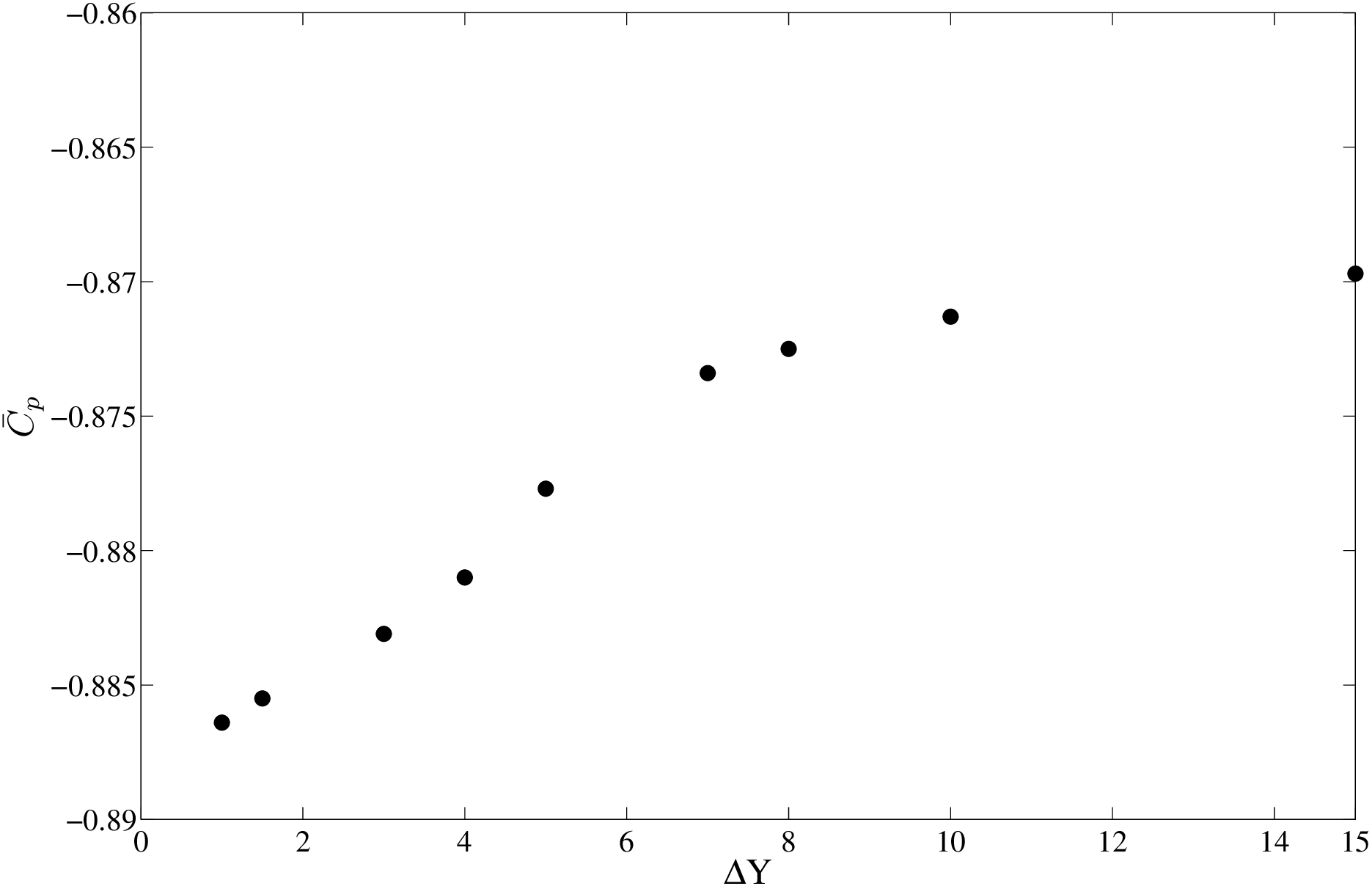}
    }
    \\
    \subfigure[Viscous force coefficient]{
      \includegraphics[scale=0.25]{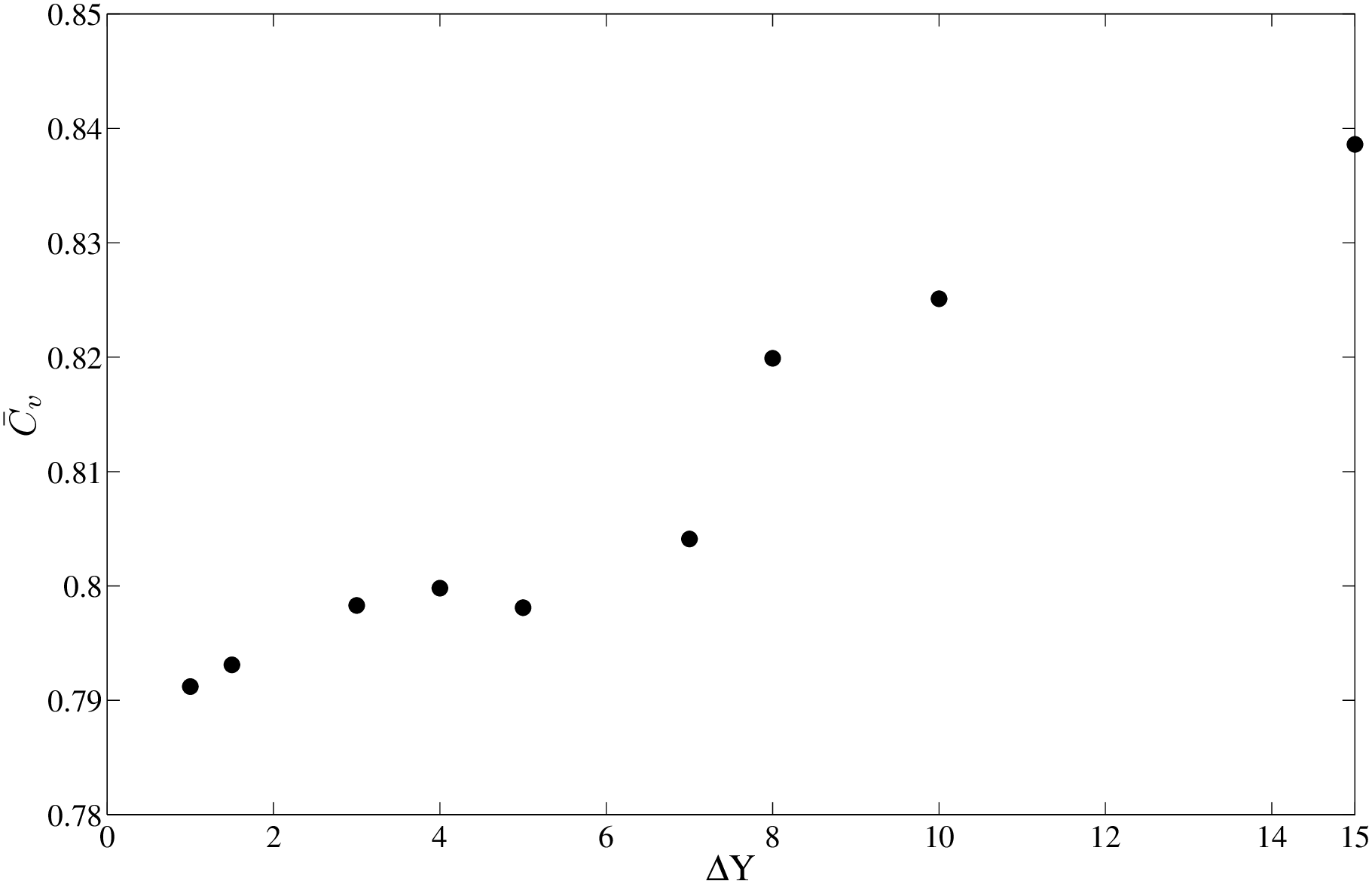}
    }
    \caption{$x$ component of mean force coefficients versus lateral separation $\Delta Y$ for double fish schooling with phase difference $\phi = \pi$.}
    \label{fig:2Fish_Phi1p0pi_Cx_DY}
  \end{figure}

  The effects of separation distance $\Delta Y$ are explored by fixing the phase difference $\phi$ equal to $\pi$, which corresponds to a mirroring between the upper and lower fish.  Fig.~\ref{fig:FishShapeChange} presents the shape variation for this symmetrical configuration in one whole period.  The symmetry ensures that both fishes experience identical force in the $x$ direction.  The relationships of the $x$ component of the mean force coefficient $\overline{C}_{x}$, and the individual contributions of pressure and viscous forces, with separation distance $\Delta Y$ are shown in Fig.~\ref{fig:2Fish_Phi1p0pi_Cx_DY}.  The overall thrust decreases monotonically, but in a complex manner, as distance is increased.  In other words,  as separation distance decreases in this symmetrical arrangement, each fish gains more benefit from the other's presence.  This benefit is achieved even at large distances, up to around 10 body lengths.  This is significantly farther than found by \citet{DongGJ:1j}, who only saw thrust influence up to around a single body length.  However, that study focused on an infinite lateral array of fish and at a higher Reynolds number (5000), both of which might affect the interaction distance.
  
When $\Delta Y$ is less than $4$, the total force coefficient reveals a nearly linear trend with respect to the separation distance $\Delta Y$.  Beyond $\Delta Y = 4$, the trend becomes more complicated and the force finally approaches to the asymptotic value $-0.03$, which is the mean net force in the solo fish.  Thus, as expected, the interaction between upper and lower fish is reduced to zero as separation increases.  It should be noted that the thrust improvement is only shown here in the time-averaged sense.  When total force is decomposed into the pressure and viscous force, the pressure force coefficient shows a relatively simpler trend which smoothly increases, while the viscous force coefficient demonstrates a complex trend which is similar to the total force coefficient.   

\begin{figure}[t]
    \centering
    \includegraphics[scale=0.45]{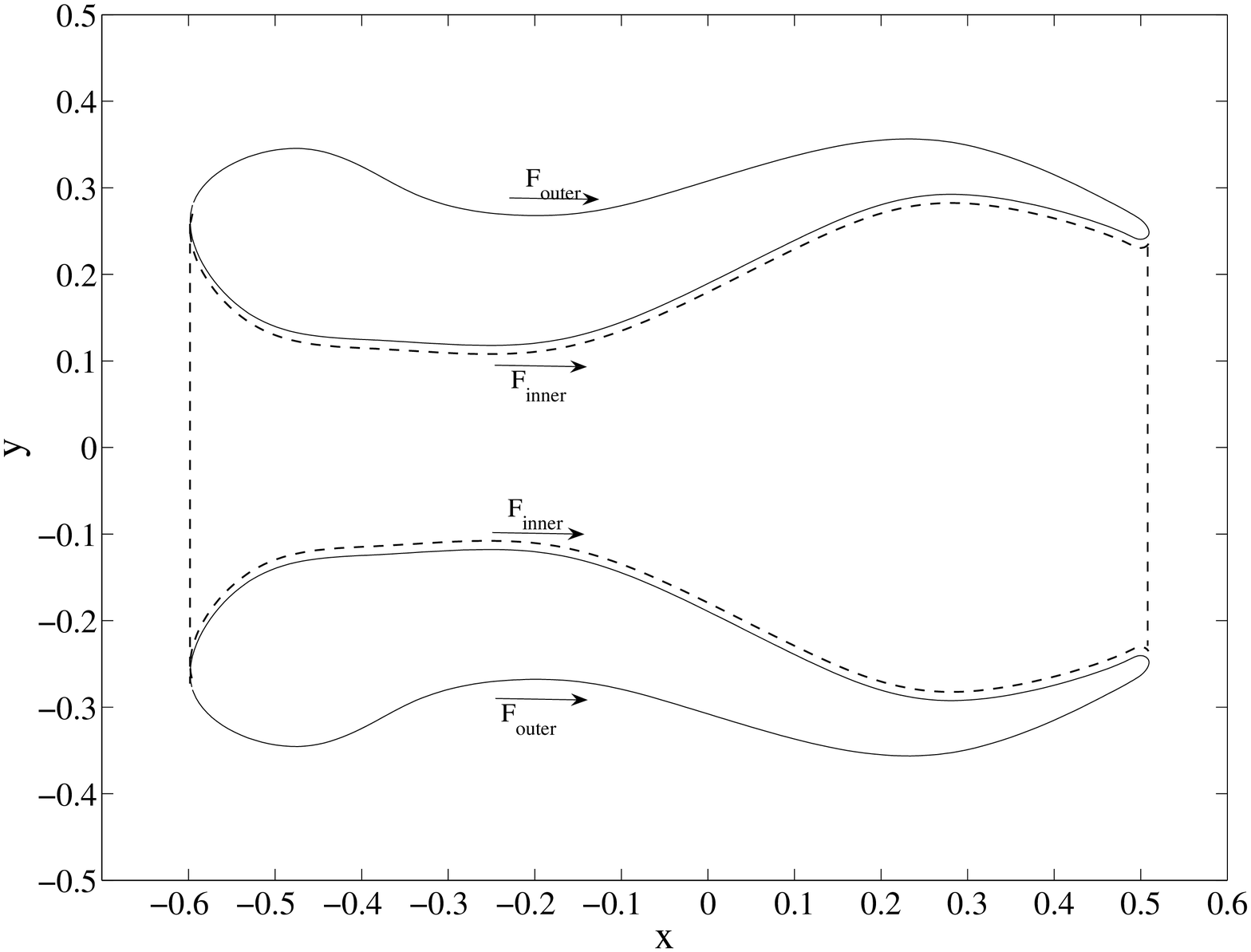}
    \caption{Schematic of control volume drawn in double fish schooling with phase difference $\phi = \pi$.}
    \label{fig:ControlV}
  \end{figure}

  In order to further understand the thrust improvement mechanisms achieved in lateral double fish schooling, we construct a control volume which follows the bottom surface of the upper fish and the top surface of the lower fish, then closes with two lateral lines from head-to-head and tail-to-tail, shown in Fig.~\ref{fig:ControlV}.  The net increase of $x$-momentum flux through the control volume is given in a time-averaged sense by
  \begin{equation}
    \int_{out} \rho\overline{u^{2}}\,d y - \int_{in} \rho\overline{u^{2}}\,d y = \overline{F}_{inner},
  \end{equation}
  where density of the fluid $\rho$ is set to be unity, and $\overline{F}_{inner}$ is the mean force acting on the inner surfaces of the fish.  Since the force on the outer portions of the fish are likely less affected by the other fish, changes in thrust would be determined mainly by changes in $\overline{F}_{inner}$.  Table~\ref{tab:u2DY} reveals that the difference of the time-averaged momentum flux normalized by separation distance between the entrance and the exit of the control volume, $\Delta\langle\overline{u^{2}} \rangle = \left(\int_{out}\overline{u^{2}}\,d y - \int_{in}\overline{u^{2}}\,d y \right)/\Delta Y$ remains almost the same as $\Delta Y$ is increased from $1$ to $4$.  Thus, the nearly linear increase of $\overline{C}_{x}$ with $\Delta Y$ can be explained by the linear dependence of net momentum increase inside the gap on $\Delta Y$.  When separation distance increases above $4$, the net momentum increase depends on $\Delta Y$ in more complex fashion, which is consistent with the complicated behavior of total mean force coefficient. 
 
  \begin{table}
    \centering
    \begin{tabular}{|c|c|}
      \hline
      $\Delta Y$ & $\Delta\langle\overline{u^{2}}\rangle$ \\
      \hline
      1.0 & 0.0706 \\
      \hline
      1.5 & 0.0707 \\
      \hline
      3.0 & 0.0705 \\
      \hline
      4.0 & 0.0706 \\
      \hline
      15.0 & 0.0627 \\
      \hline
    \end{tabular}
    \caption{Normalized Time-averaged increase in momentum flux versus separation distance}
    \label{tab:u2DY}
  \end{table}
  
  We postulate that at small separation distance $\Delta Y$, the gap between the fish pair behaves like a peristaltic pump \citep{Jaffrin:1j}, which also has a transverse wave traveling from the entrance to the exit.  Such a pumping mechanism, which delivers a jet of fluid through the exit of the gap, is responsible for producing large net thrust at small separations.  However, when separation distance between the upper and lower fish is large, more complex effects from the head and tail and nonlinear wake interactions appear.  Therefore, more thrust can be generated in smaller separation distance by the peristaltic pumping mechanism and comparatively less thrust is obtained at larger distance, which combines the peristaltic pumping mechanism with wake interactions. 

\begin{figure}[t]
    \centering
    \subfigure[$\Delta Y = 1.5$]{
      \includegraphics[scale=0.35]{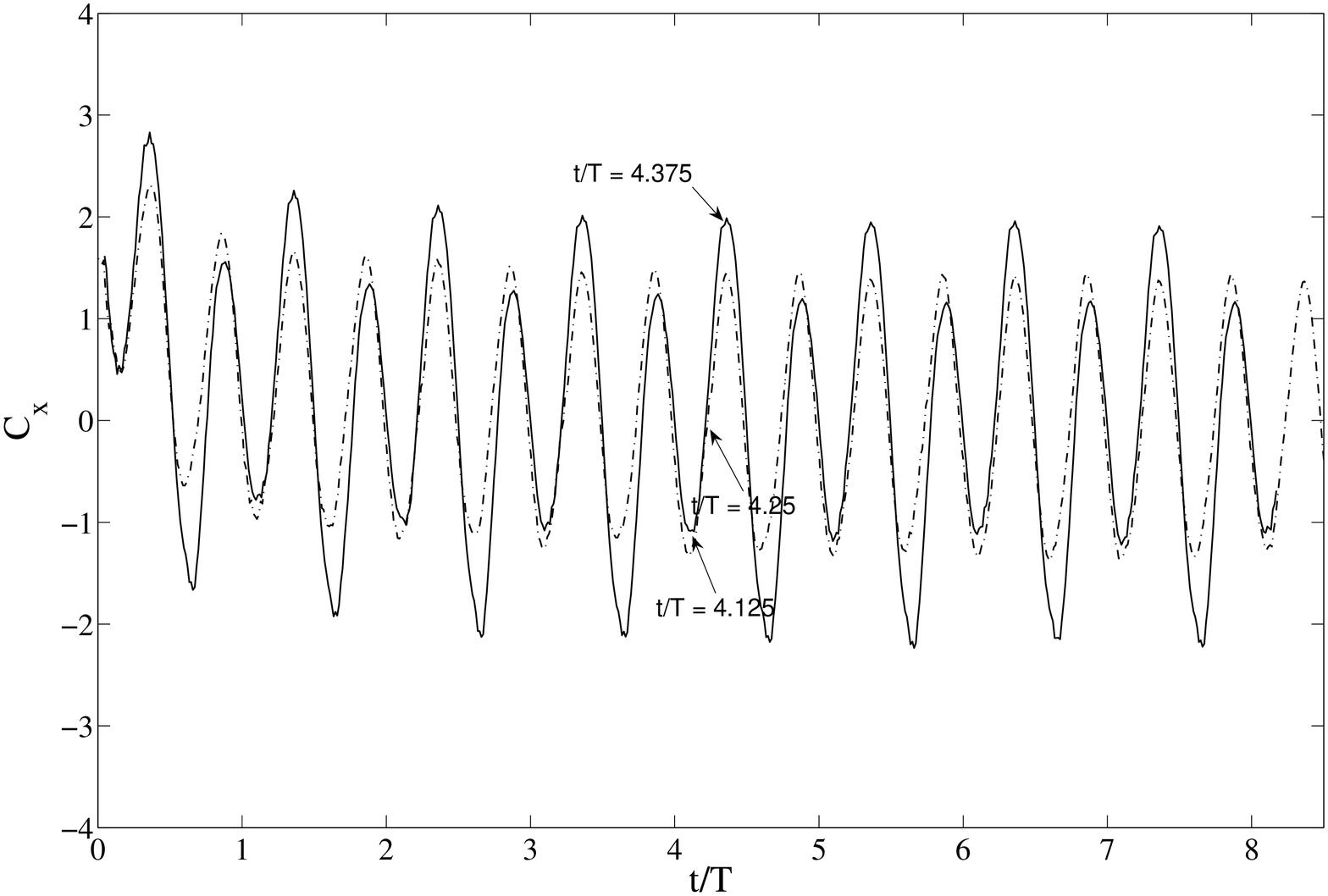}
    }
    \subfigure[$\Delta Y = 5.0$]{
      \includegraphics[scale=0.35]{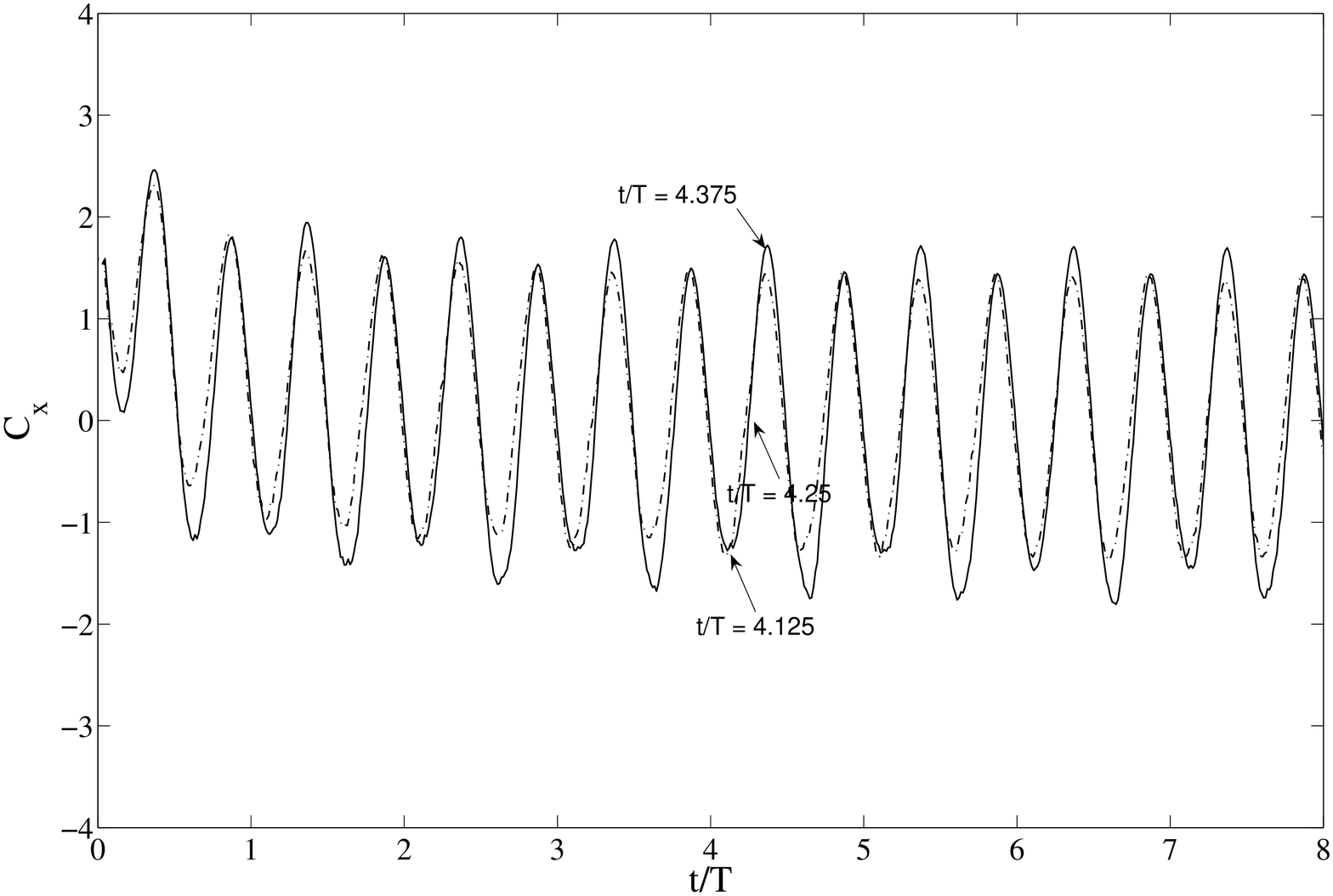}
    }
    \caption{Time-varying $x$ component of force coefficient in the case of double fish schooling with $\phi = \pi$. $-$ represents double fish, $-.$ represents solo fish.}
    \label{fig:2Fish_DY1p5_Phi1p0pi_Cx}
  \end{figure}

  An example of the resulting time histories of the force coefficient in the $x$ direction, $C_{x}$, is shown in Fig.~\ref{fig:2Fish_DY1p5_Phi1p0pi_Cx} for two cases, $\Delta Y = 1.5$ and $\Delta Y = 5.0$.  Compared with the force history of the solo fish, the presence of the other fish results in a more complicated history, with notably different values for the peaks in each period.  However, as $\Delta Y$ increases, the force coefficient history becomes more like the solo fish history, with a sinusoidal shape, since the interaction between upper and lower fish is reduced to almost zero.  
  
\begin{figure}[t]
    \centering
    \subfigure[Pressure distribution]{
      \includegraphics[scale=0.35]{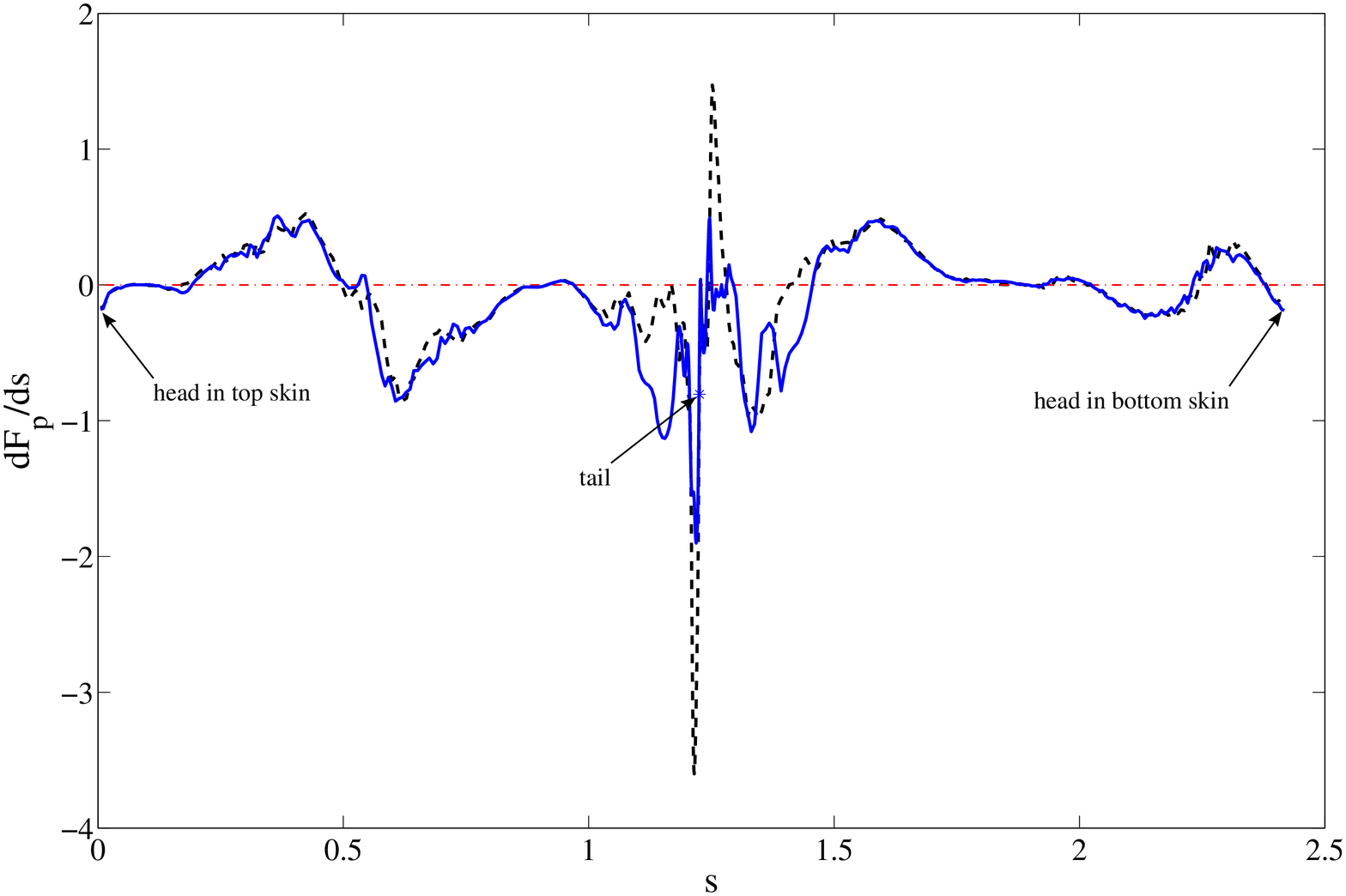}
    }
    \subfigure[Regions of force generation]{
      \includegraphics[scale=0.35]{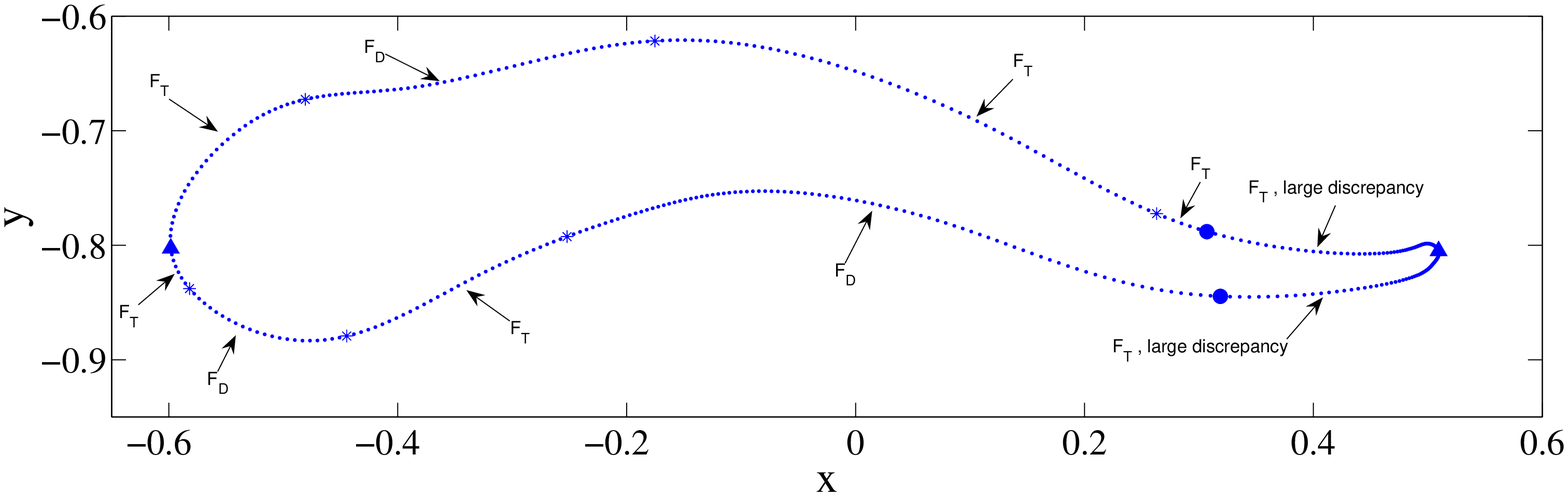}
    }
    \caption{(a) Distribution of pressure contribution to the $x$ component of force at $t/T = 4.125$ and $\phi = \pi$. $--$, $\Delta Y = 1.5$, $-$, $\Delta Y = 5.0$, $-.$, zero line. (b) Regions of notable pressure at $t/T = 4.125$. $*$, represents critical point; $\bigtriangleup$, represents head and tail; tail region between $\bullet$ represents the region of significant force variation.}
    \label{fig:FpalongS_Phi1p0pi_T3p3}
  \end{figure}

\begin{figure}[t]
    \centering
    \includegraphics[scale=0.45]{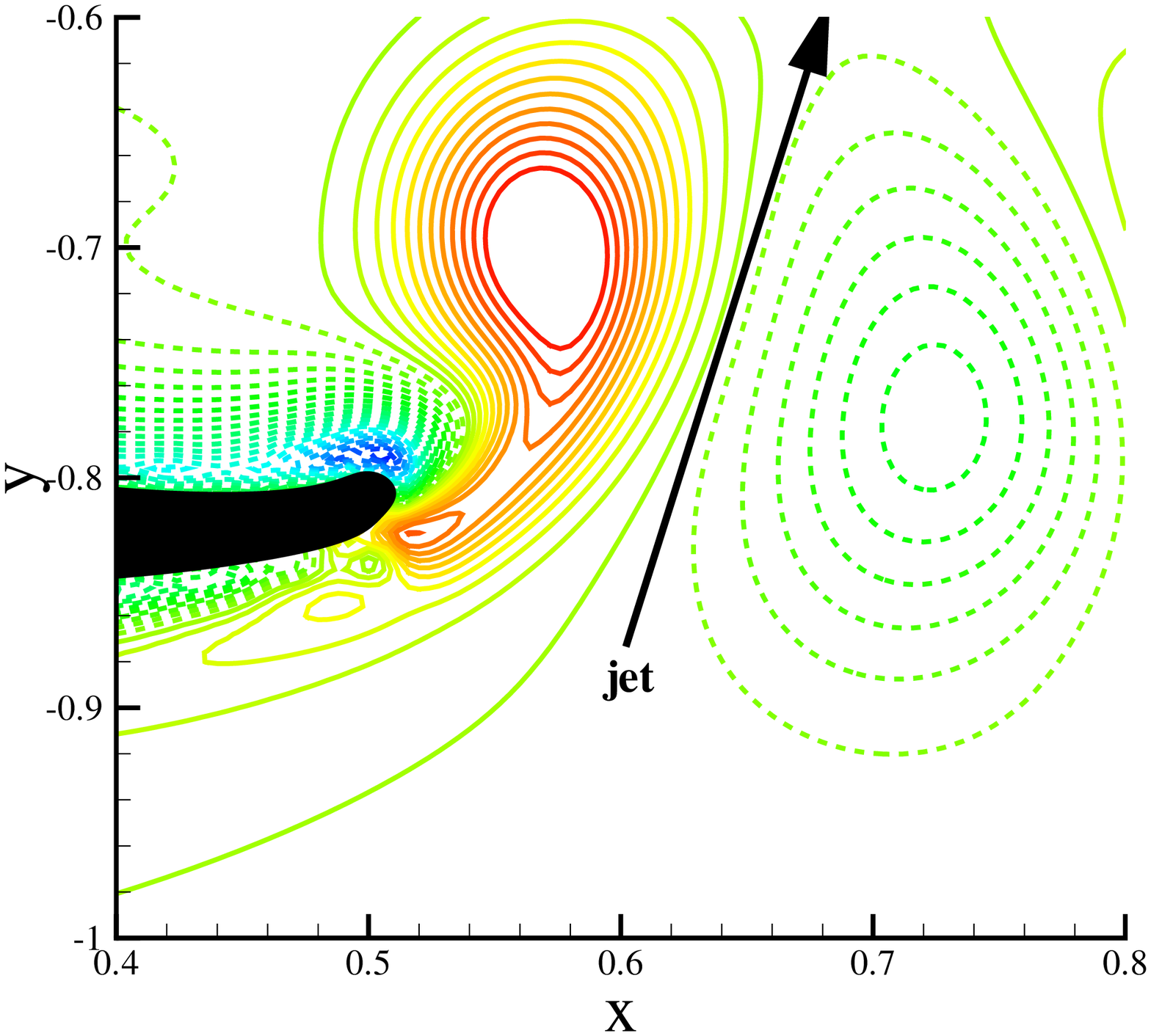}
    \caption{Vorticity field in tail vicinity at $t/T = 4.125$, dashed lines correspond to negative vorticity.}
    \label{fig:ZoomVortTail_Phi1p0pi_T3p3}
  \end{figure}

\begin{figure}[t]
    \centering
    \subfigure[Pressure distribution]{
      \includegraphics[scale=0.35]{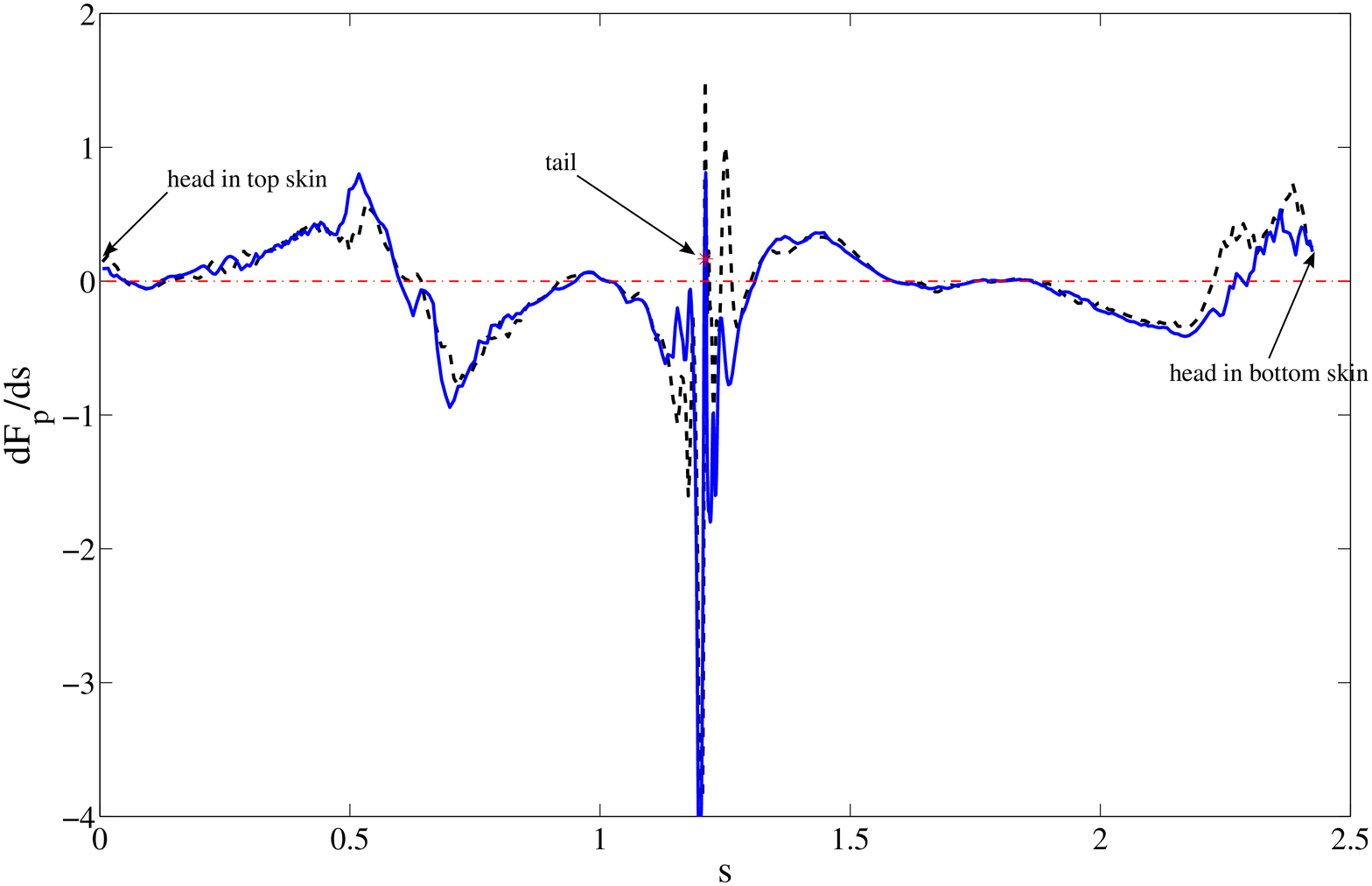}
    }
    \subfigure[Regions of force generation]{
      \includegraphics[scale=0.35]{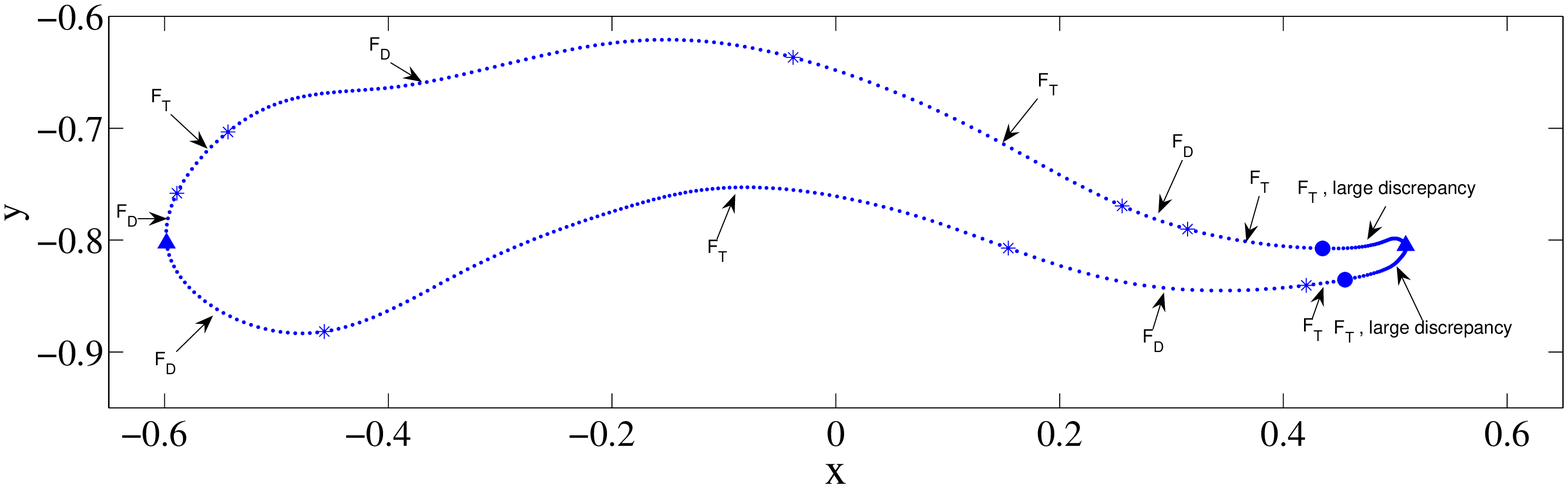}
    }
    \caption{(a) Distribution of pressure contribution to the $x$ component of force at $t/T = 4.25$ and $\phi = \pi$. $--$, $\Delta Y = 1.5$, $-$, $\Delta Y = 5.0$, $-.$, zero line. (b) Regions of notable pressure at $t/T = 4.125$. $*$, represents critical point; $\bigtriangleup$, represents head and tail; tail region between $\bullet$ represents the region of significant force variation.}
    \label{fig:FpalongS_Phi1p0pi_T3p4}
  \end{figure}

\begin{figure}[t]
    \centering
    \includegraphics[scale=0.45]{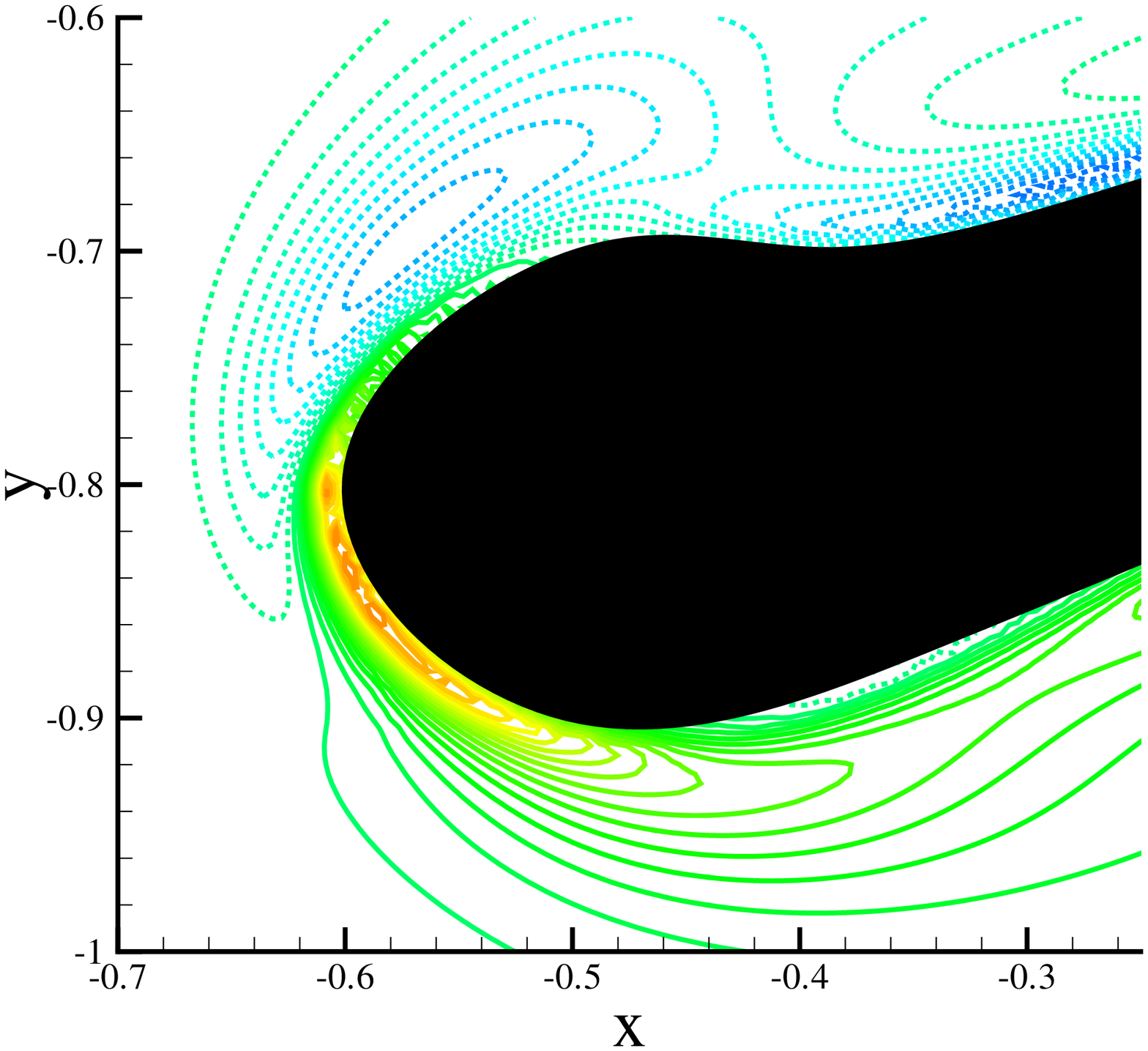}
    \caption{Vorticity field in head vicinity at $t/T = 4.25$, dashed lines correspond to negative vorticity.}
    \label{fig:ZoomVortHead_Phi1p0pi_T3p4}
  \end{figure}

\begin{figure}[t]
    \centering
    \includegraphics[scale=0.45]{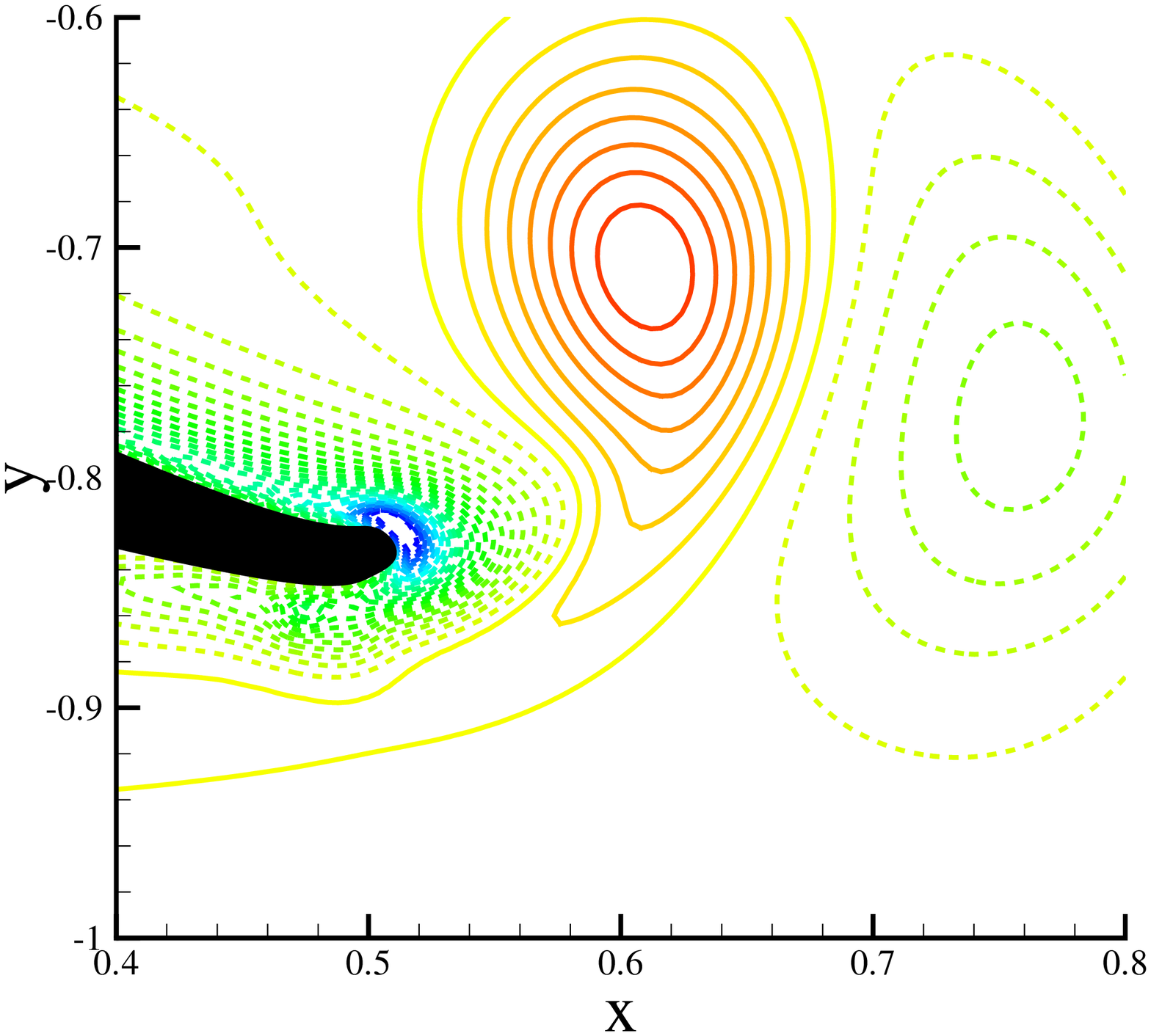}
    \caption{Vorticity field in tail vicinity at $t/T = 4.25$, dashed lines correspond to negative vorticity.}
    \label{fig:ZoomVortTail_Phi1p0pi_T3p4}
  \end{figure}

\begin{figure}[t]
    \centering
    \subfigure[Pressure distribution]{
      \includegraphics[scale=0.35]{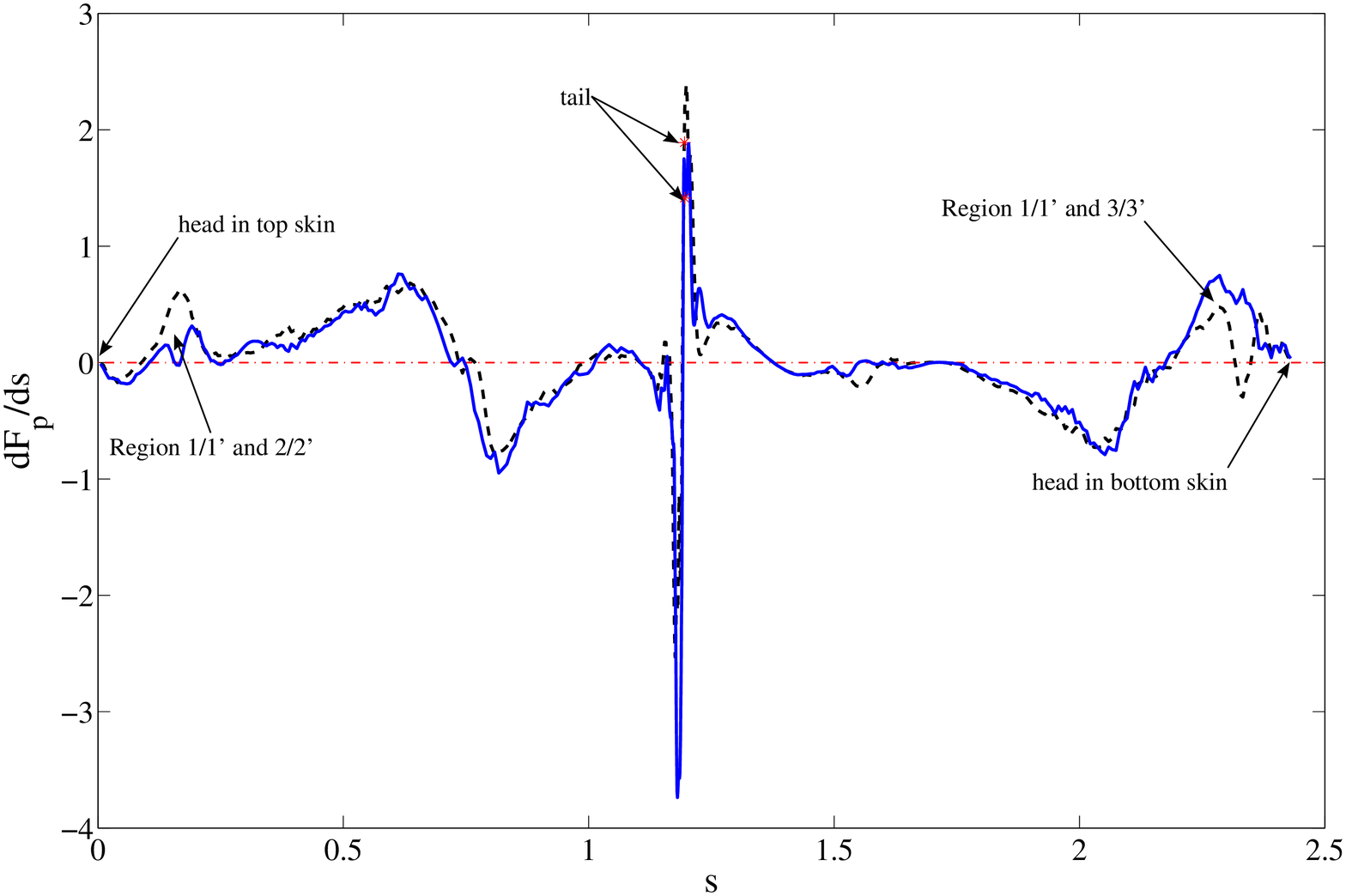}
    }
    \subfigure[Regions of force generation]{
      \includegraphics[scale=0.35]{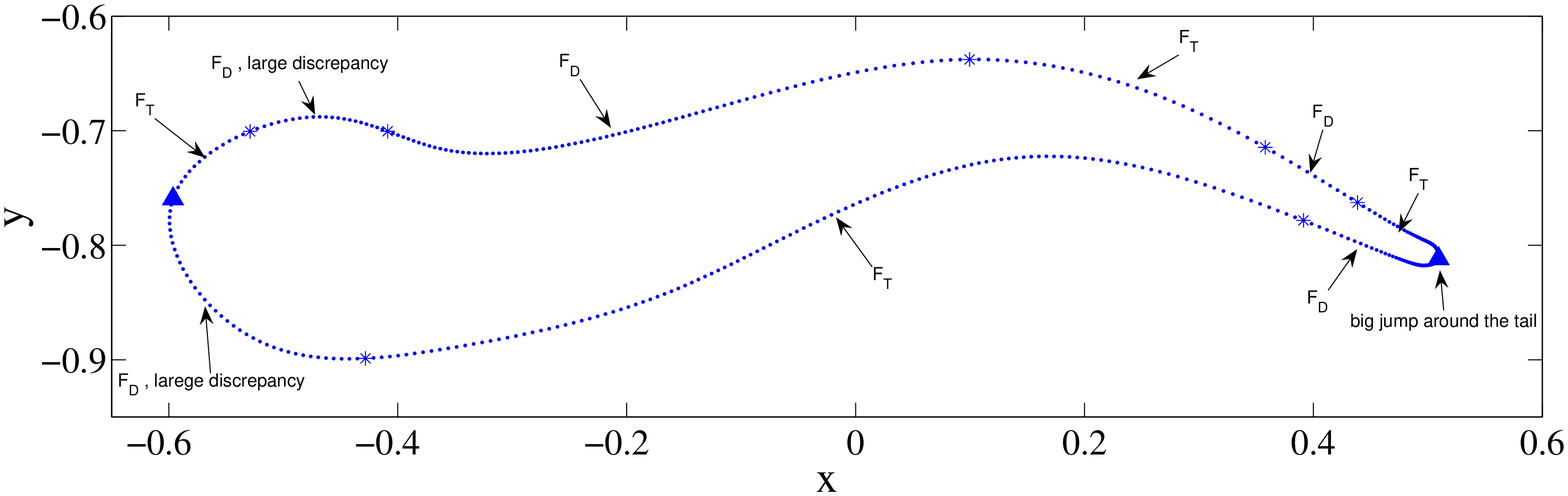}
    }
    \caption{(a) Distribution of pressure contribution to the $x$ component of force at $t/T = 4.375$ and $\phi = \pi$. $--$, $\Delta Y = 1.5$, $-$, $\Delta Y = 5.0$, $-.$, zero line. (b) Regions of notable pressure at $t/T = 4.125$. $*$, represents critical point; $\bigtriangleup$, represents head and tail; tail region between $\bullet$ represents the region of significant force variation.}
    \label{fig:FpalongS_Phi1p0pi_T3p5}
  \end{figure}

\begin{figure}[t]
    \centering
    \subfigure[$\Delta Y = 1.5$]{
      \includegraphics[scale=0.45]{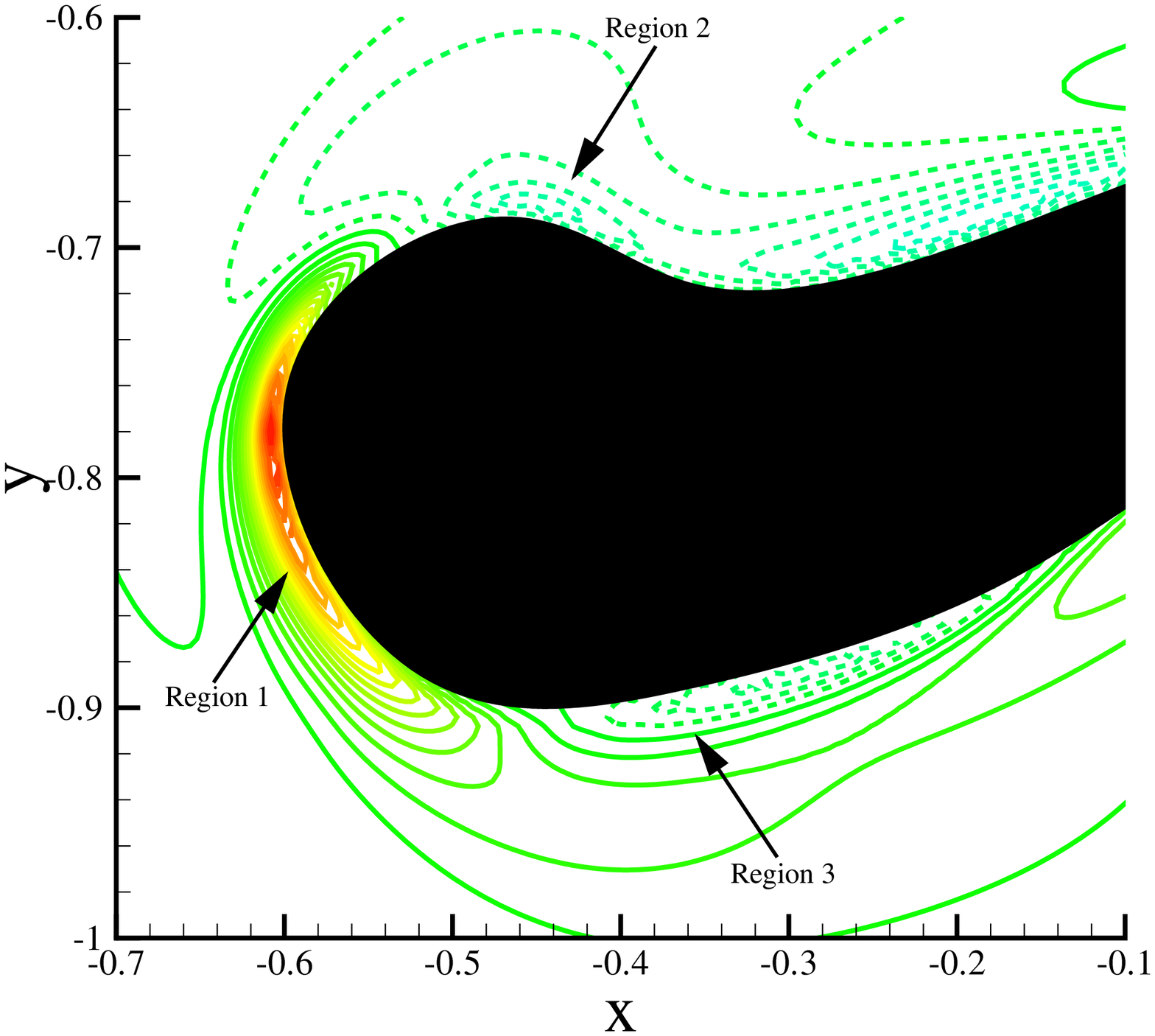}
    }
    \subfigure[$\Delta Y = 5.0$]{
      \includegraphics[scale=0.45]{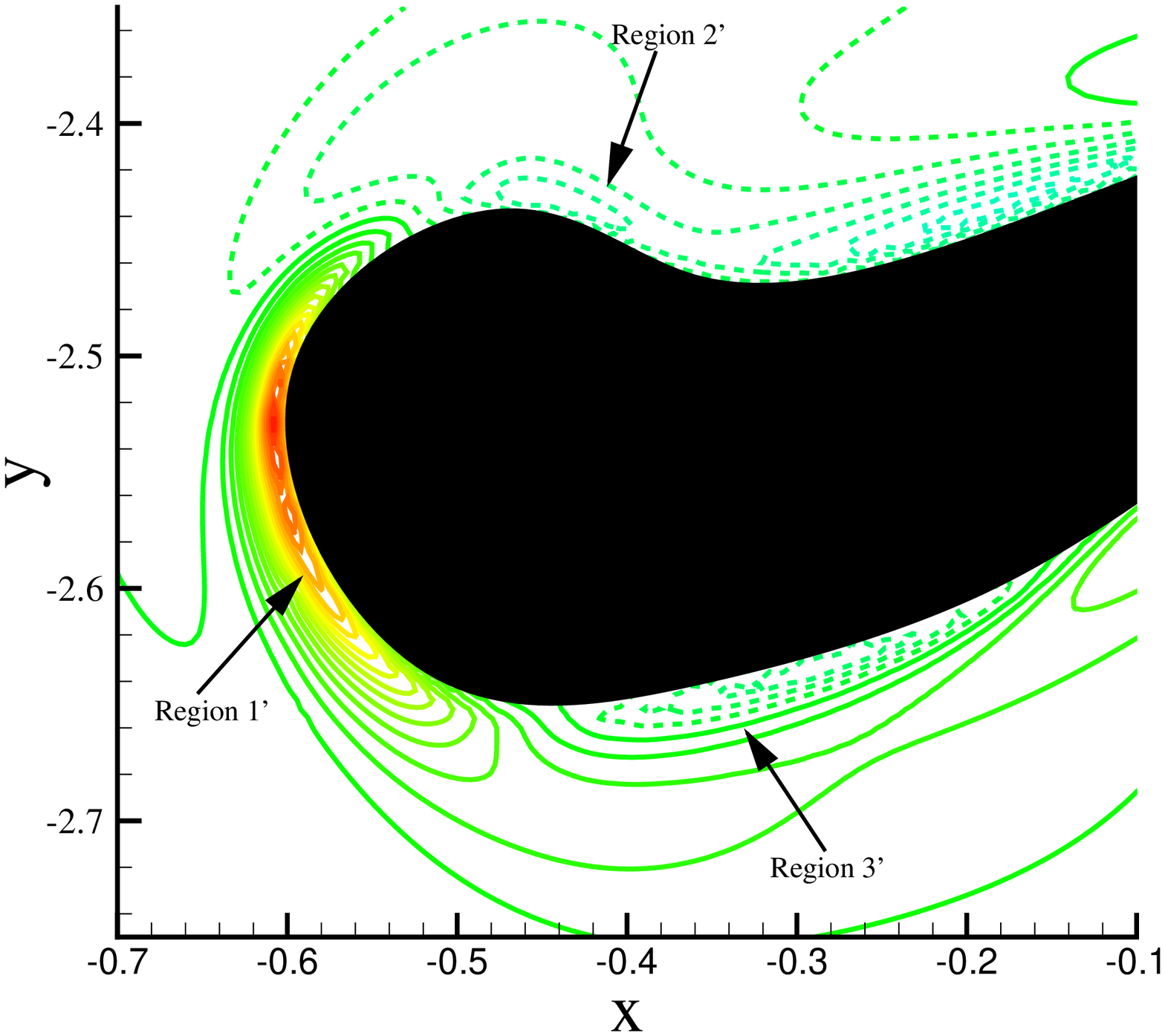}
    }
    \caption{Vorticity field in head vicinity at $t/T = 4.375$ and $\phi = \pi$, dashed lines correspond to negative vorticity.}
    \label{fig:ZoomVortHead_Phi1p0pi_T3p5}
  \end{figure}

\begin{figure}[t]
    \centering
    \includegraphics[scale=0.45]{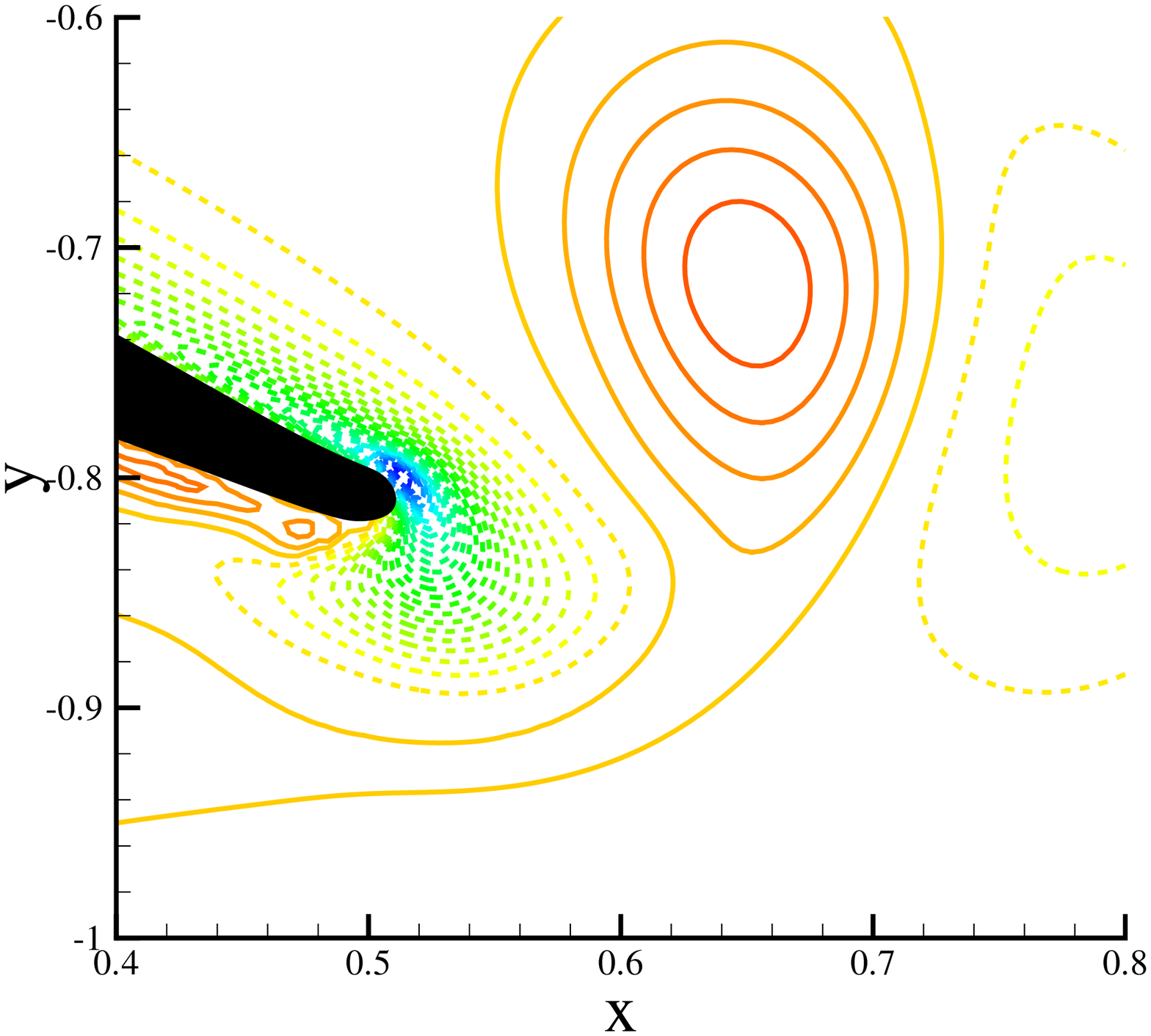}
    \caption{Vorticity field in tail vicinity at $t/T = 4.375$ and $\phi = \pi$, dashed lines correspond to negative vorticity.}
    \label{fig:ZoomVortTail_Phi1p0pi_T3p5}
  \end{figure}

  It is informative to probe the notable differences in force exhibited in Fig.~\ref{fig:2Fish_DY1p5_Phi1p0pi_Cx} by exploring the force distribution on the body surface.  Three representative instants of the lower fish are shown from Figs.~\ref{fig:FpalongS_Phi1p0pi_T3p3} to \ref{fig:ZoomVortTail_Phi1p0pi_T3p5}: $t/T = 4.125$ corresponds to the lower peak or maximum thrust, $t/T = 4.25$ corresponds to almost zero force, and $t/T = 4.375$ corresponds to the peak force, or maximum drag.  Both pressure force distribution and local vorticity field are shown in these figures. 

  Since the pressure force coefficient increases smoothly as $\Delta Y$ increases, we focus our attention on two separation distances, $\Delta Y = 1.5$ and $\Delta Y = 5.0$.  We note on the fish profile plots in this series, a critical region in which a large discrepancy is found in the pressure force distribution between $\Delta Y = 1.5$ and $\Delta Y = 5.0$.  When $t/T = 4.125$, the critical region that is represented by the portion between the solid circles in Fig.~\ref{fig:FpalongS_Phi1p0pi_T3p3} (b) is located close to the fish tail and accounts for about $21\%$ of surface perimeter.  The force on the remaining portions of the fish surface in these two cases remains approximately matched.  Segments of the fish surface experiencing thrust $F_{T}$ and drag $F_{D}$ are also denoted in Fig.~\ref{fig:FpalongS_Phi1p0pi_T3p3} (b).  The pressure force at $t/T = 4.125$ on each segment is consistent with the fish shape and motion at this instant.  The local vorticity field near the tail shown in Fig.~\ref{fig:ZoomVortTail_Phi1p0pi_T3p3} reveals that the trailing edge vortex has been nearly shed off the tail at this instant.  This vortex which forms a jet with the previously-shed vortex in the wake, as indicated on the figure.  This process has a significant influence on the pressure force exerted on the tail.

  When $t/T = 4.25$ in Fig.~\ref{fig:FpalongS_Phi1p0pi_T3p4}, the critical region is also located close to the tail, but accounts for about $8\%$ of surface perimeter which is significantly smaller than the one in $t/T = 4.125$.  Simultaneously, in the remaining portion, the pressure force in two cases $\Delta Y = 1.5$ and $\Delta Y = 5.0$ is roughly matched except the area around the head on the outer side of the fish.  At this instant, the thrust and the drag on all segments are consistent with the instantaneous relative motion of the surface and fluid, and the instantaneous fish shape.  In particular, the thrust between $s = 0.5$ and $s = 1$ is generated by the rearward motion of the lateral bulge on the inner surface.  The vorticity field close to the head, shown in Fig.~\ref{fig:ZoomVortHead_Phi1p0pi_T3p4} reveals vorticity layers of alternating sign by side-by-side motion of the nose.  These layers are subtly modified by the presence of the other fish, resulting in some discrepancy in force distribution as separation distance changes.  The tail, shown in Fig.~\ref{fig:ZoomVortTail_Phi1p0pi_T3p4}, is currently in upward motion, having shed the previous vortex and generating a new trailing edge vortex.  The associated jet is significantly weaker at this instant, and not strong enough to produce net thrust.  This process is slightly modified by the other fish, as is evident in the pressure distribution of Fig.~\ref{fig:FpalongS_Phi1p0pi_T3p4}(a). 

  When $t/T = 4.375$, Fig.~\ref{fig:FpalongS_Phi1p0pi_T3p5} shows that the critical region is confined to the tail, but there are notable differences around the head, on the inner and outer surfaces, which is also subtly revealed in the vorticity field close to the head in Fig.~\ref{fig:ZoomVortHead_Phi1p0pi_T3p5}.  Regions of notable difference are labeled on this figure.  The vorticity close to the tail in Fig.~\ref{fig:ZoomVortTail_Phi1p0pi_T3p5} exhibits the starting point of a newly growing vortex and the lingering vortex produced in the previous half-beat.  At this instant, the new trailing edge vortex has not been generated yet, and the previous vortex produced in the last half-beat has been transported farther from the tail.  Thus, no jet is observed in this case, which results in maximum drag. 

  It is known from Fig.~\ref{fig:2Fish_DY1p5_Phi1p0pi_Cx} that from $t/T = 4.125$ to $t/T = 4.375$, the net force changes from thrust to drag.  It is also noted that the tail acts as a significant source of thrust and that the critical region in which the tail's contribution is important is decreasing in size in this interval.  This decrease in size is related to the process of vorticity generation at the tail.  When $t/T = 4.375$, this region reaches to the minimum size, associated with maximum drag.  Though the pressure force distribution on the surface presents the same trend for  $\Delta Y = 1.5$ and $\Delta Y = 5.0$, the total pressure force is smaller in $\Delta Y = 5.0$, as shown in Fig.~\ref{fig:2Fish_Phi1p0pi_Cx_DY} (b).  The presence of the other fish contributes a small but important effect, primarily at the tail, and, to a lesser extent, the head.  

\begin{figure}[t]
    \centering
    \includegraphics[scale=0.45]{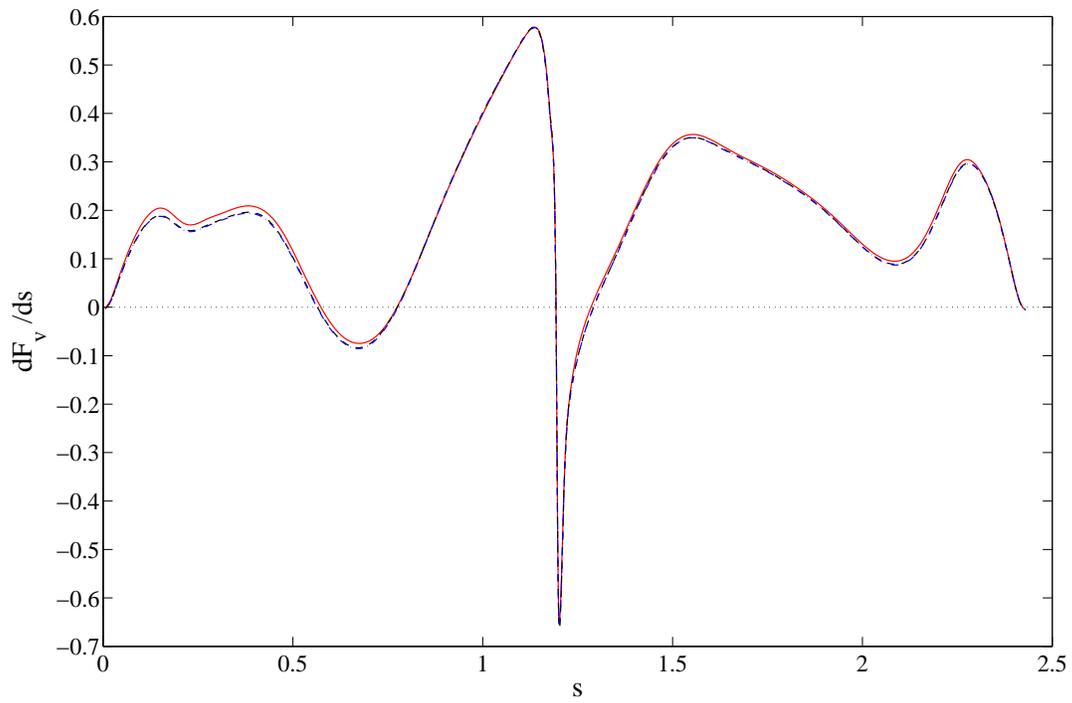}
    \caption{Distribution of viscous contribution to the $x$ component of force along body surface at $t/T = 4.375$ and $\phi = \pi$. $-.$, $\Delta Y = 1.5$; $--$, $\Delta Y = 5.0$; $-$, $\Delta Y = 8.0$; $..$, zero line.}
    \label{fig:FvalongS_Phi1p0pi_T3p5}
  \end{figure}

  The viscous force distribution on the lower fish is examined for three different separation distances at one instant, associated with the maximum drag in Fig.~\ref{fig:FvalongS_Phi1p0pi_T3p5}.  It is found that a small but significant discrepancy exists on the inner surface near the head in the lower fish, which is the surface exposed to the gap between the two fishes.  This discrepancy is primarily between $\Delta Y = 5.0$ and $\Delta Y = 8.0$, while there is almost no discernible difference between $\Delta Y = 1.5$ and $5.0$, which is consistent with the trend of mean viscous force found in Fig.~\ref{fig:2Fish_Phi1p0pi_Cx_DY}(c).   Thus for small gap sizes, the presence of another fish has a notable effect on the viscous drag, but this effect is insensitive to changes in separation distance.  For separation larger than $\Delta Y = 4.0$, the viscous drag transitions to an asymptotic trend toward the solo fish value.  Although the trailing edge vorticity close to the tail remains the same sign, as evident from Fig.~\ref{fig:ZoomVortTail_Phi1p0pi_T3p5}, the $x$ component of the tangential vector changes the sign around the tail, which causes the $x$ component of the viscous force to change the sign as well.  Hence, the instantaneous vorticity in the vicinity of the tail acts as a significant source of both viscous drag and thrust. 

\begin{figure}[t]
    \centering
    \subfigure[$t/T = 7.375$]{
      \includegraphics[scale=0.45]{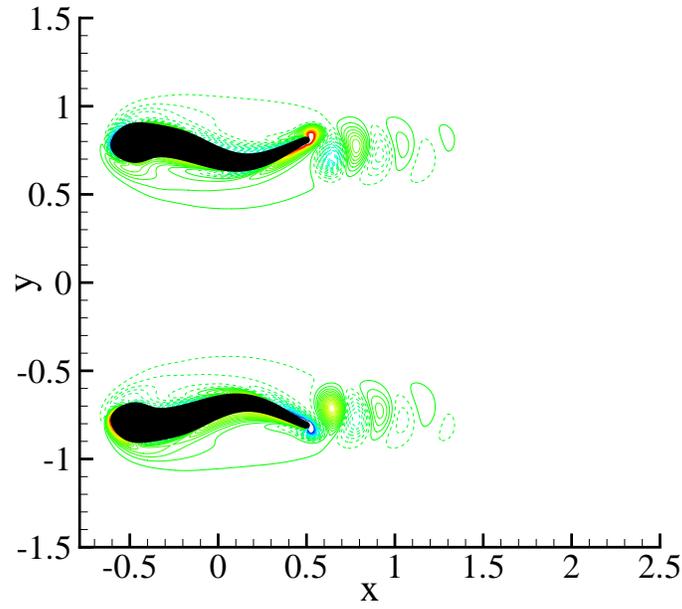}
    }
    \subfigure[$t/T = 7.625$]{
      \includegraphics[scale=0.45]{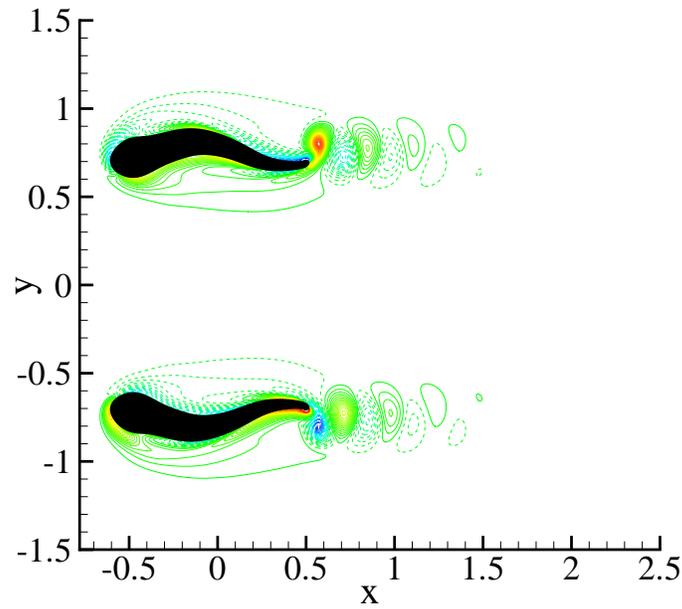}
    }
    \caption{Vorticity contours of lateral pair at $\Delta Y = 1.5$ and $\phi = \pi$}
    \label{fig:2Fish_DY1p5_Phi1p0pi_Vort}
  \end{figure}

  The resulting vorticity contours at different instants are shown in Fig.~\ref{fig:2Fish_DY1p5_Phi1p0pi_Vort} for $\Delta Y = 1.5$.  The first corresponds to an instant of the maximum drag and the other corresponds to the maximum thrust.  Since the thrust improvement remains in a small amount, the vorticity wake reveals a vortex shedding pattern between a classical von K\'arm\'an vortex street and the inversed von K\'arm\'an vortex street.  This is consistent with previous results, such as by \citet{DongGJ:1j}.  

\begin{figure}[t]
    \centering
    \includegraphics[scale=0.55]{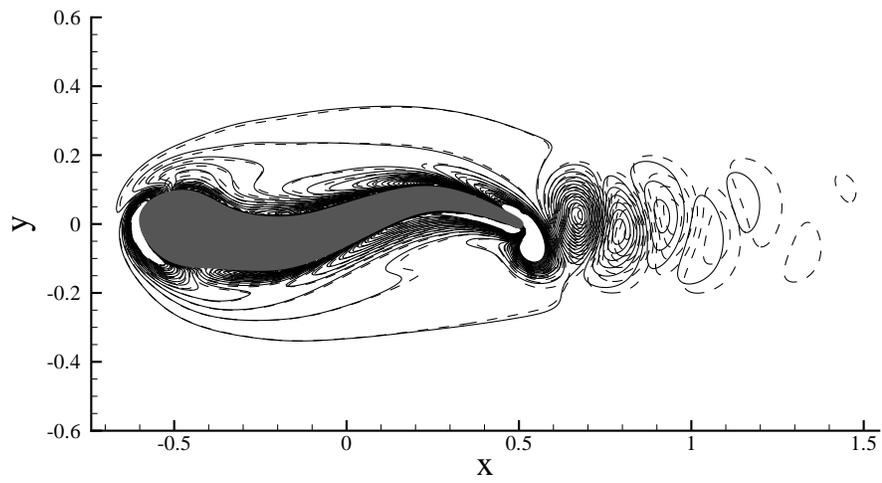}
    \caption{Comparison of vorticity wake between solo fish and lateral double fish schooling at $t/T = 7.5$. $-$, solo fish; $--$, lateral fish schooling.}
    \label{fig:Compare2FishSoloFish_StrT6p0}
  \end{figure}

  Fig.~\ref{fig:Compare2FishSoloFish_StrT6p0} compares the vorticity between the solo fish and the lower of the two fishes separated by $\Delta Y = 1.5$, at the same instant in the cycle.  This figure illustrates that the vorticity pattern has been significantly modified due to the presence of the other fish.  The modification is most striking in the wake, which would be expected to have an important influence on the force exerted on the tail, in light of the relationship elucidated in this section.  In particular, this modification is consistent with the discrepancies in pressure force distribution near the tail as separation distance changes.    

  \subsubsection{Effects of phase difference $\phi$}
  
\begin{figure}[t]
    \centering
    \includegraphics[scale=0.45]{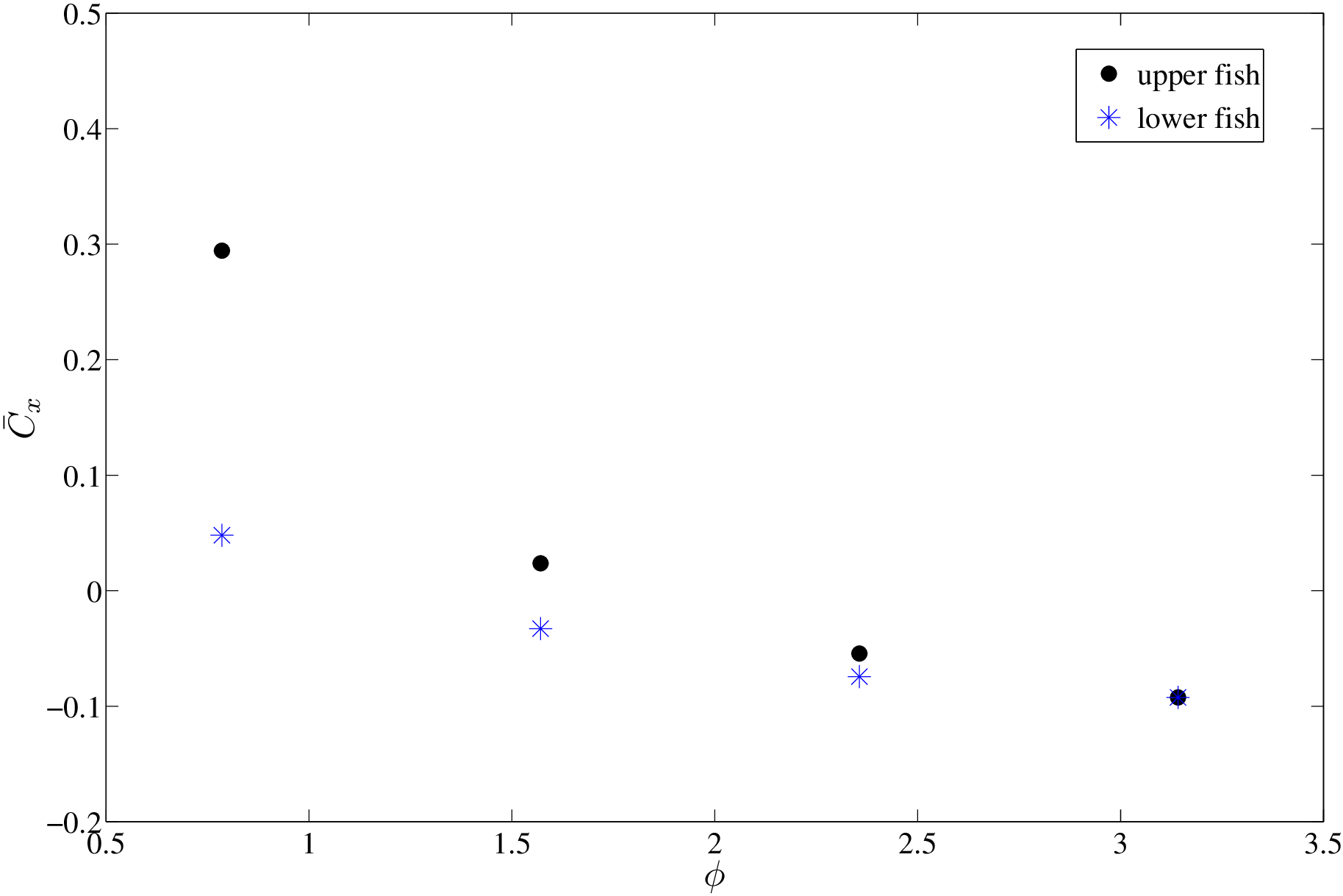}
    \caption{$x$ component of mean force coefficient versus $\phi$ for double fish schooling with $\Delta Y = 1.5$.}
    \label{fig:2Fish_Cx_Phi}
  \end{figure}

  Fixing the lateral separation distance at $\Delta Y = 1.5$ and Strouhal number at $St = 0.8$, phase difference $\phi$ is varied from $\pi/4$ to $\pi$ with interval $\pi/4$.  The mean force coefficient $\overline{C}_{x}$ versus phase difference $\phi$ for both fishes is presented in Fig.~\ref{fig:2Fish_Cx_Phi}.  Both upper and lower fish present a smoothly declining trend of total force coefficient as $\phi$ is increased, until $\phi = \pi$ when both upper and lower fish reach to the largest mean thrust.  In contrast, $\phi = \pi/4$ corresponds to maximum net drag for both fishes.  In addition, the lower fish achieves more benefit in terms of thrust, while only the symmetrical case $\phi = \pi$ provides the same and maximum thrust in both two fish.  

\begin{figure}[t]
    \centering
    \includegraphics[scale=0.45]{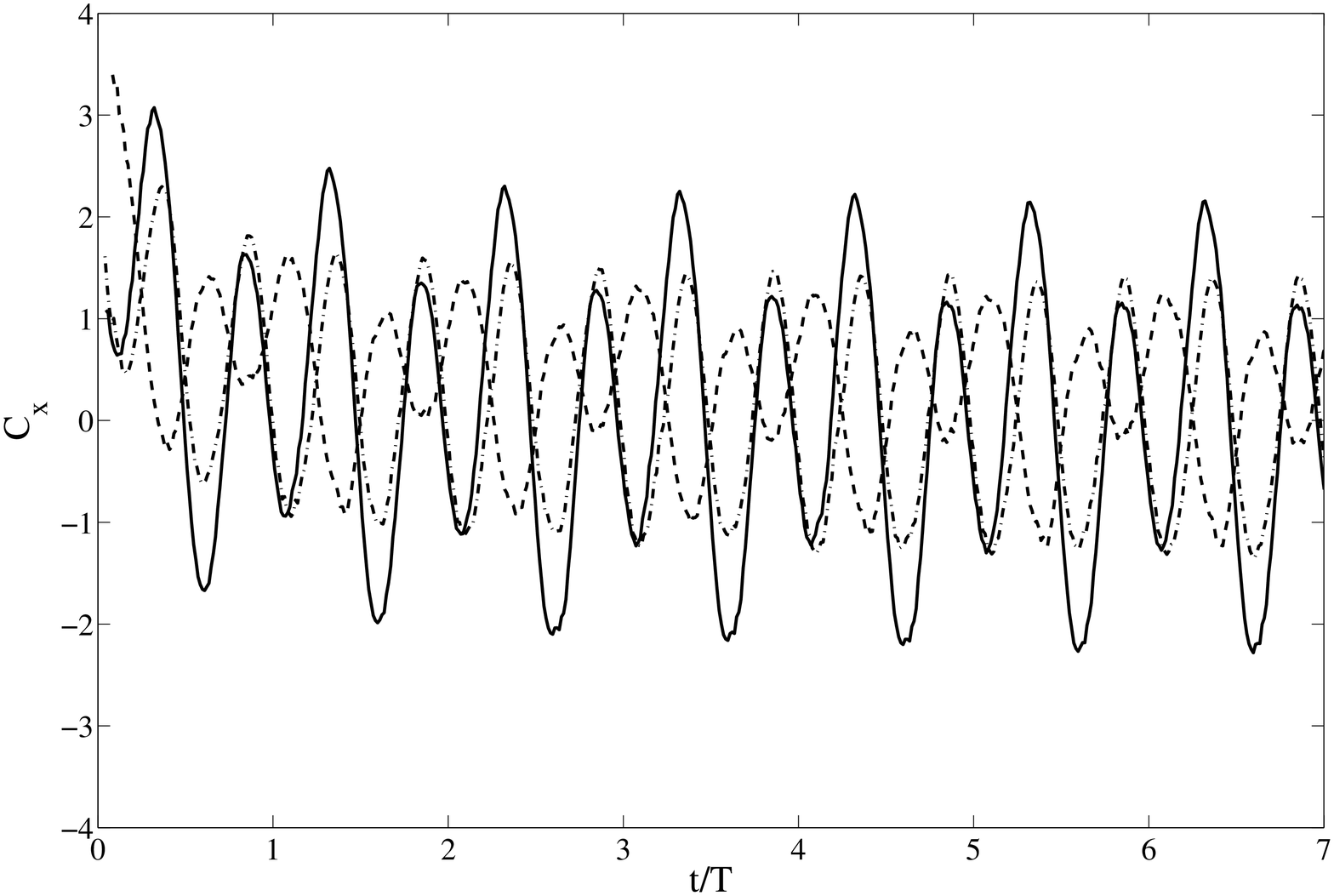}
    \caption{Time-varying $x$ component of force coefficient in the case of double fish schooling with $\Delta Y = 1.5$ and $\phi = \pi/2$. $--$, represents upper fish; $-$, represents lower fish; $-.$, represents solo fish.}
    \label{fig:2Fish_DY1p5_Phi0p5pi_Cx}
  \end{figure}

  An example of the time-varying $x$ component of the force coefficients for upper and lower fish with $\phi = \pi/2$ is shown in Fig.~\ref{fig:2Fish_DY1p5_Phi0p5pi_Cx}.  It is interesting to note that the force on the lower fish is nearly identical to that on the solo fish for some portion of each period, but significantly different for the remaining portion.  In other words, the fish is only affected by the other fish for a fraction of each period.
  
\begin{figure}[t]
    \centering
    \includegraphics[scale=0.45]{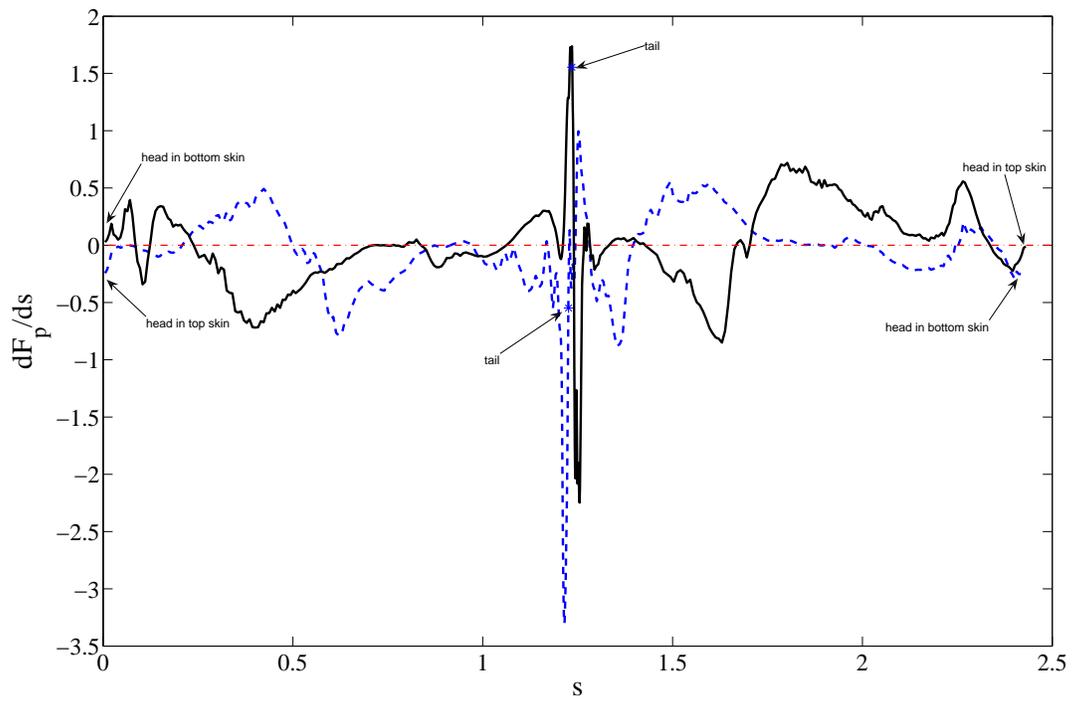}
    \caption{Distribution of pressure contribution to the $x$ component of force at $t/T = 4.125$ with $\phi = \pi/2$ and $\Delta Y = 1.5$. $--$, represents lower fish; $-$, represents upper fish; $-.$, represents zero line.}
    \label{fig:FpalongS_Phi0p5pi_T3p3}
  \end{figure}

\begin{figure}[t]
    \centering
    \subfigure[Upper fish]{
      \includegraphics[scale=0.35]{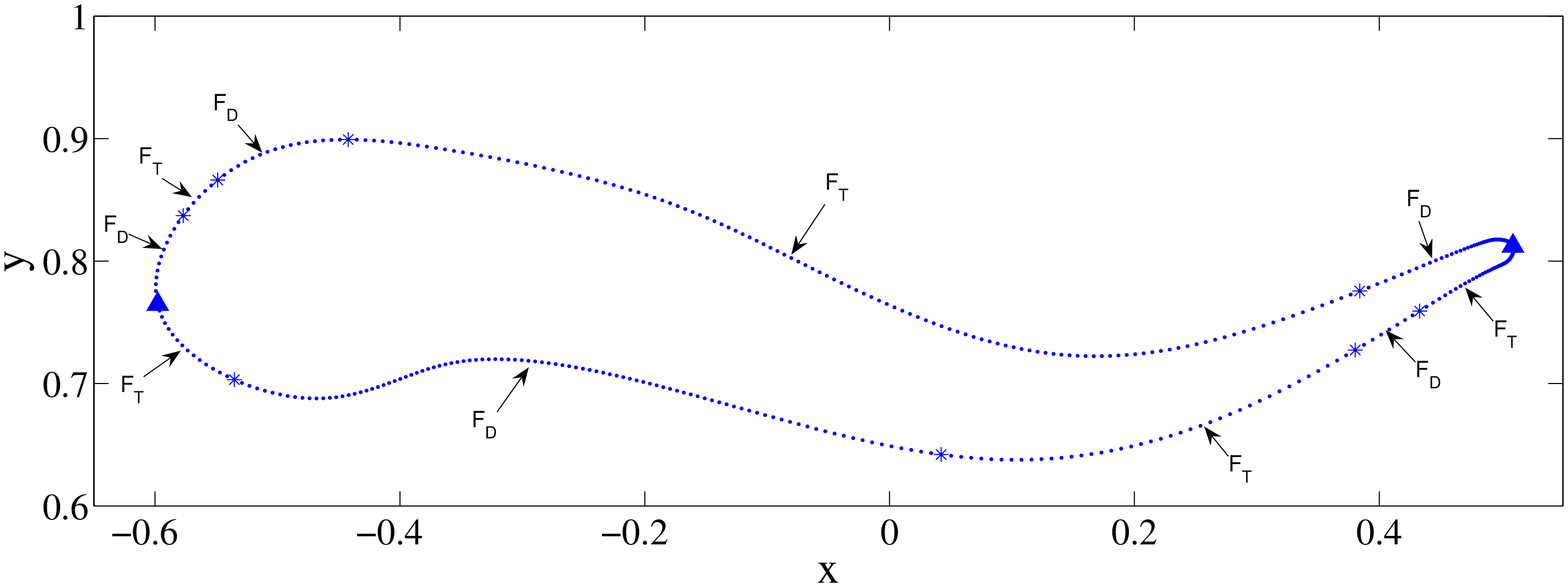}
    }
    \subfigure[Lower fish]{
      \includegraphics[scale=0.35]{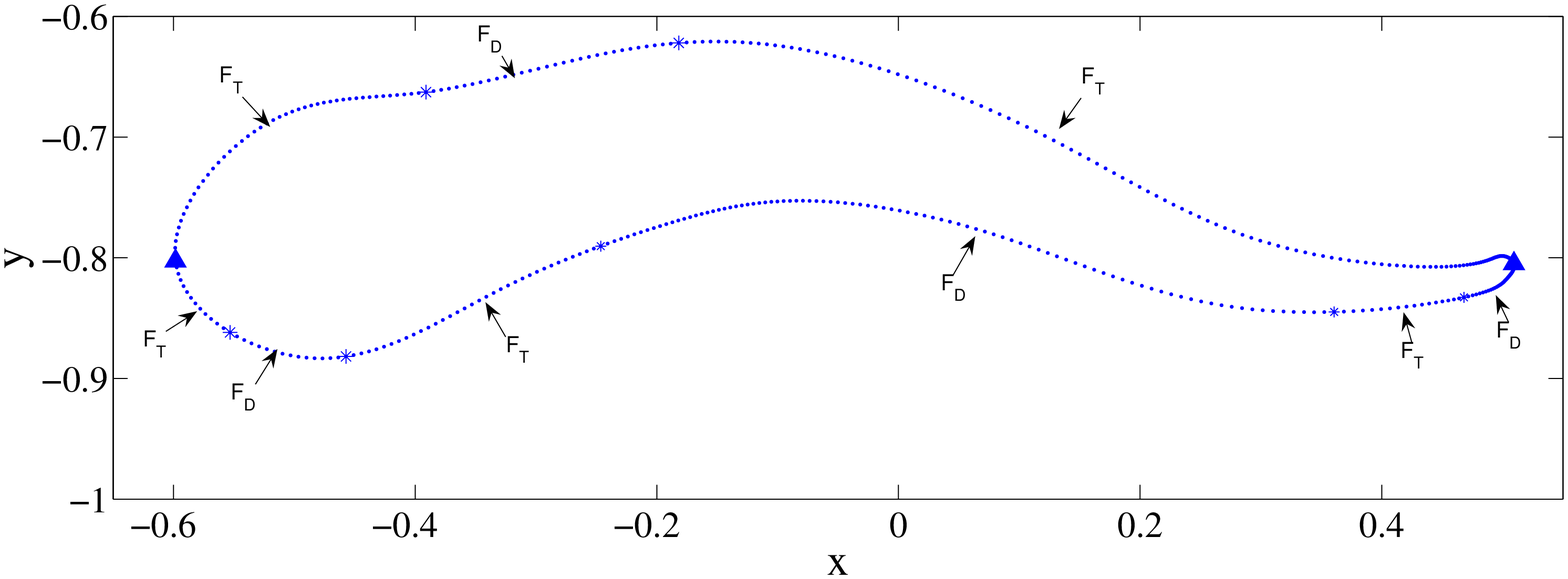}
    }
    \caption{Regions of notable pressure at $t/T = 4.125$ with $\phi = \pi/2$ and $\Delta Y = 1.5$. $*$, represents critical point; $\bigtriangleup$, represents head and tail.}
    \label{fig:ULpanles_SegsForce_Phi0p5pi_T3p3}
  \end{figure}

  In order to further demonstrate the force generation mechanism in this asymmetrical lateral schooling,  we choose one extreme instant $t/T = 4.125$ for the phase difference $\pi/2$, at which the lower fish has maximum thrust while the upper fish has maximum drag.  Fig.~\ref{fig:FpalongS_Phi0p5pi_T3p3} shows the distribution of pressure contribution to the $x$ component of force on the upper and lower fish and Fig.~\ref{fig:ULpanles_SegsForce_Phi0p5pi_T3p3} reveals notable pressure regions on each fish.  The pressure force illustrates completely different distributions in the lower and upper fish.  In particular, the tail of the lower fish is generating a net thrust, while the force in the vicinity of the tail of the upper fish is almost zero.  Hence, at this instant, the lower fish obtains more benefits in thrust from the pairing.  The thrust and drag regions on upper and lower fish are consistent with the instantaneous motions and shapes of each fish, as well.  

  \subsection{Lateral schooling of three fish}
  
\begin{figure}[t]
    \centering
    \includegraphics[scale=0.45]{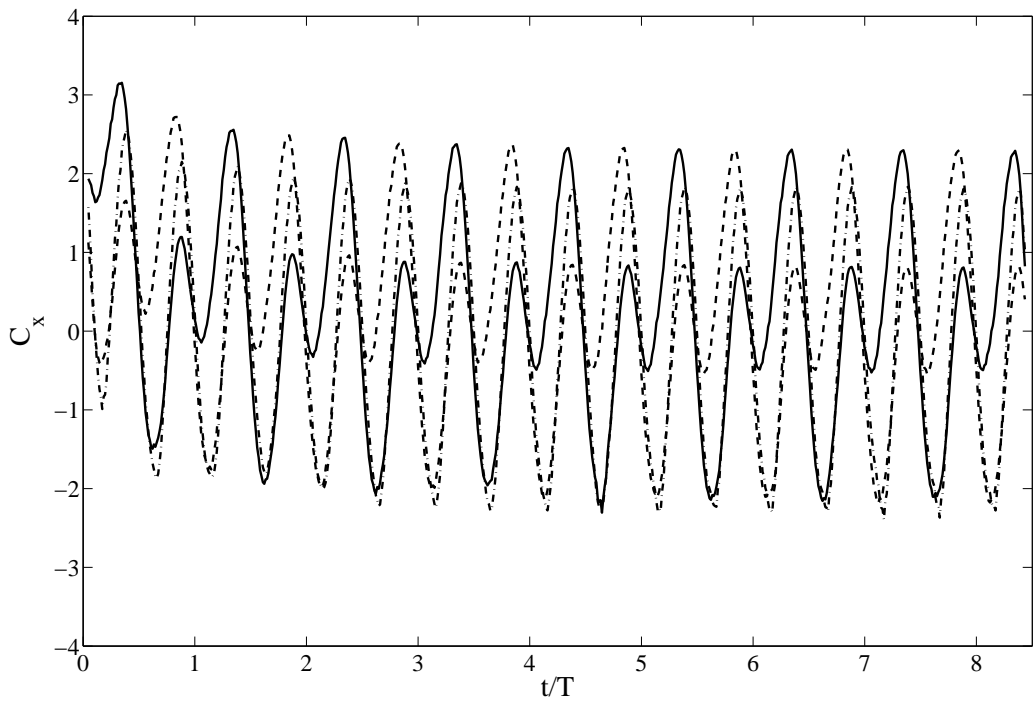}
    \caption{Time-varying $x$ component of force coefficient in the case of three fish schooling with $\Delta Y = 1.5$ and $\phi = \pi$. $--$, presents upper fish; $-$, presents lower fish; $-.$, presents center fish.}
    \label{fig:3Fish_Cx}
  \end{figure}

\begin{figure}
    \centering
    \subfigure[$t/T = 4.15$]{
      \includegraphics[scale=0.5]{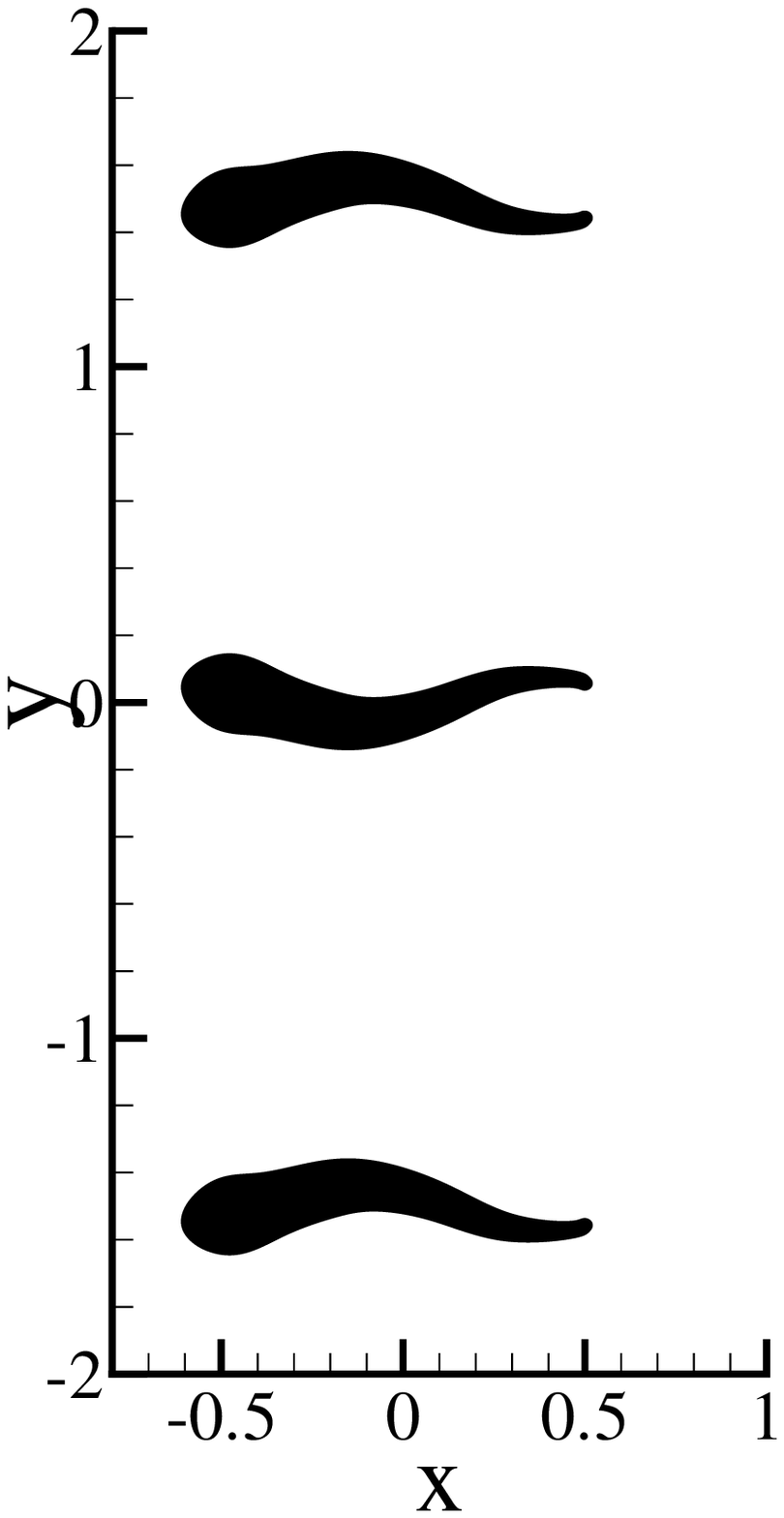}
    }
    \subfigure[$t/T = 4.65$]{
      \includegraphics[scale=0.5]{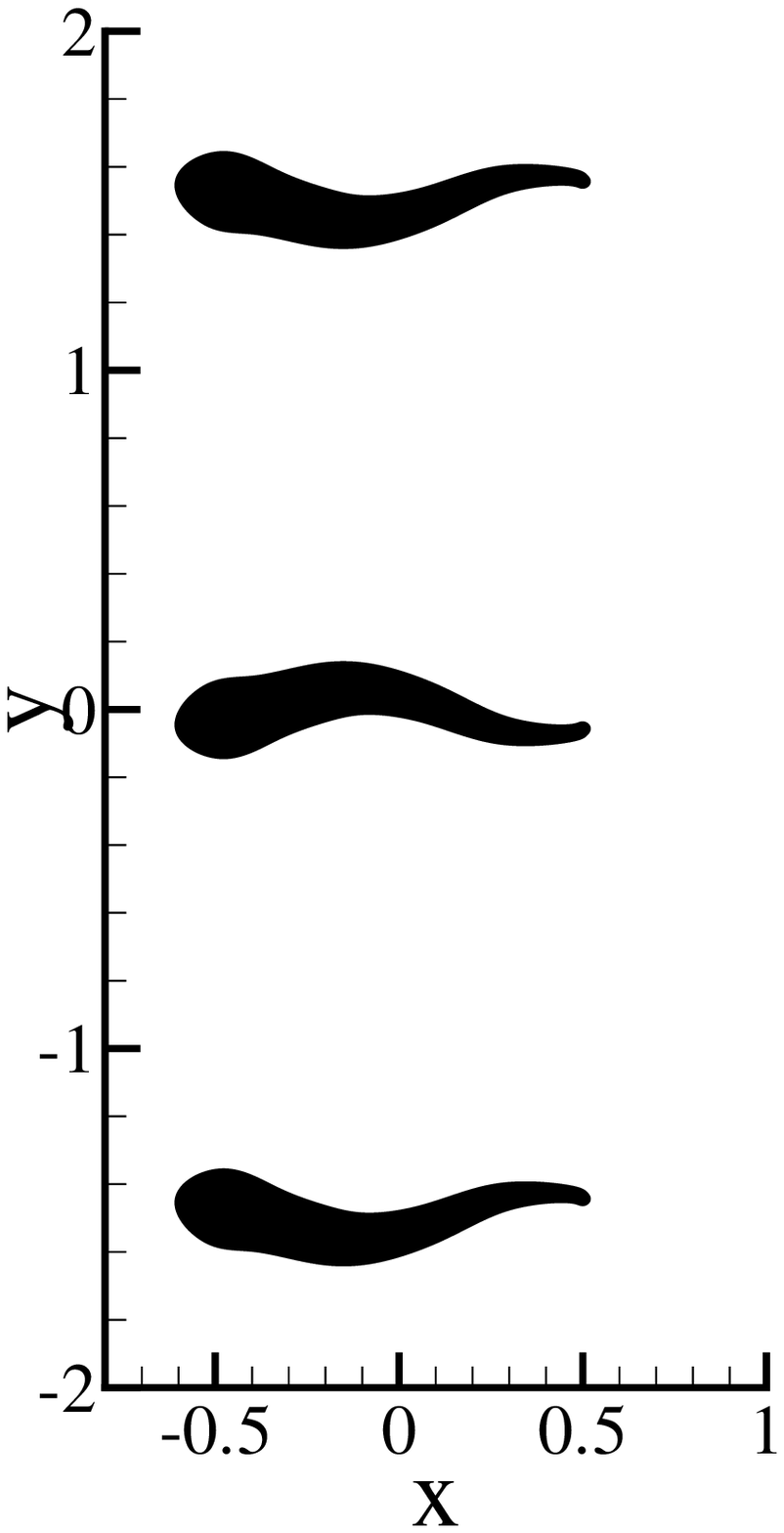}
    }
    \caption{Layout of lateral three fish with $\Delta Y = 1.5$ and $\phi = \pi$.}
    \label{fig:3ShapeTs}
  \end{figure}

  In this section, we examine a lateral configuration of three fishes, in the arrangement shown in Fig.~\ref{fig:FishSketchMap}.  In this case, we restrict our attention to the same phase difference ($\phi = \pi$) of the center fish with respect to either the lower fish or the upper fish.  In other words, two symmetrical pairs are constructed: one consists of the center and upper fish, and the other one consists of the center and lower fish.  Compared with the solo fish, only the center fish obtains thrust improvement, while both upper and lower fish experience mean net drag, which is shown in the force histories of Fig.~\ref{fig:3Fish_Cx}.  The mean value of force coefficient on the center fish is $-0.25$ and the mean value on the lower and upper fish is approximately matched, $0.16$.  The time-varying force coefficient of the center fish is nearly sinusoidally varying.  Due to the balanced interactions between its upper and lower side, the force on the center fish is not biased toward either side, which explains the sinusoidal history.  The force histories on the upper and the lower fish more closely resemble the `two-peak' histories of the fish pairs from the previous section.  However, it is important to note that each pair in this three-fish arrangement is not identical to its two-fish counterpart.  In other words, each pair is influenced by the presence of the third, more remote fish.  It is also interesting to note that the lower peak (thrust) of the force coefficient in the center fish is alternately aligned with one of the lower and upper fish.  The fish configurations corresponding to these matched peaks are shown in  Fig.~\ref{fig:3ShapeTs}.  The instant when the peaks of the center fish and upper fish match ($t/T = 4.15$) is nearly identical to the instant of peak thrust in the two-fish configuration (Fig.~\ref{fig:2Fish_DY1p5_Phi1p0pi_Cx}).  When the center fish is aligned with the upper fish, the upper pair becomes the dominant pair while the lower pair contributes almost zero force.  In summary, the force generation of three fishes in a lateral arrangement exhibits a behavior which is consistent with the single and double fish configurations, though with slight differences due to the presence of the third fish.
  
  \section{Conclusions}
  
  With the goal to research the hydrodynamics of lateral fish schooling, we have applied a viscous vortex particle method to one, two and three tethered fishes undergoing undulatory shape changes.  The Reynolds number was held fixed at 100, and the Strouhal number used in all lateral schooling configurations was $St = 0.8$, which was found to generate nearly zero mean net force in the $x$ direction for a solo fish.
  
  For double fish schooling, we explored the effect of separation distance $\Delta Y$ to the thrust improvement.  As $\Delta Y$ increases, the thrust augmentation in a time-averaged sense diminishes until asymptotically reaching the force generated by the solo fish, and interactions between upper and lower fish are reduced to zero.  This augmentation persists even at separation distances of 10 body lengths.  The pressure and viscous contributions are both affected.  The pressure portion, which contains a net thrust, is primarily affected at the tail by the presence of the other fish.  This interaction changes during the course of a stroke, as the vorticity shedding itself changes.  
  
  Secondly, we explored the effect of phase difference $\phi$ to the thrust improvement.  As $\phi$ increases from zero to $\pi$, the net force on both fishes declines (tending from net drag to net thrust).  But at most phase differences, the lower fish gains more benefit in terms of thrust, until $\phi = \pi$, at which both fishes experience the same thrust.  Asymmetrical configurations induced by phase difference result in asymmetrical force generation on each fish.  
  
  For lateral schooling of three fishes, only the center fish obtains thrust improvement, while both outer fishes experience a mean net drag.  Due to the balanced interactions on upper and lower side,  the force on the center fish is not biased toward either side.  However, the upper and lower pairs are both affected by the third fish.  This unfortunately seems to preclude the opportunity for superposition of results from elemental pairs to obtain results in larger lateral arrays.  We also found alternating alignment of the force on the center fish with the upper and lower fish during each stroke.  When the center fish is aligned with the upper fish, the upper pair becomes the dominant pair while the lower fish experiences almost zero force, and vice versa.  

  This study represents an opening step in a sequence of studies of hydrodynamic interactions in biolocomotion.  In future work, it will be important to study three-dimensional schools, self-propelling individuals, and larger groups with streamwise and lateral members.

  {\bf List of symbols}
 \begin{itemize}
 \itemindent = 0.5in
  \labelsep = 0.5in
    \item[$s$] backbone generating parameter
    \item[$A_{0}$] the undulation amplitude at $s = 0$
    \item[$k$] the growth rate of undulation amplitude
    \item[$f$] tail-beat frequency
    \item[$t$] time
    \item[$T$] period of undulation ($1/f$)
    \item[$A_{T}$] the peak-to-peak tail-beat amplitude
    \item[$St$] Strouhal number
    \item[$F_{x}$] $x$ component of the force
    \item[$C_{x}$] $x$ component of the force coefficient
    \item[$U_{\infty}$] the free stream velocity
    \item[$L$] fish chord length
    \item[$Re$] Reynolds number
    \item[$\Delta Y$] separation distance between adjacent fish
    \item[$\phi$] phase difference between adjacent fish
    \item[$\rho$] density of the fluid
    \item[$\Delta\langle\overline{u^{2}}\rangle$] time-averaged net momentum flux between adjacent fish
    \item[$F_{T}$] thrust
    \item[$F_{D}$] drag
    \item[$F_{inner}$] net force acting on the inner surfaces of two adjacent fish
    \end{itemize}

  \par\vspace{3ex}\par\noindent
Support for this work by the National Science Foundation, under award CBET-0645228, is gratefully acknowledged.

  \bibliographystyle {jexpbiol}

  \clearpage
  
  \clearpage

  \clearpage
  
  \clearpage
  
  \clearpage
  
  \clearpage

  \clearpage

  \clearpage

  \clearpage

  \clearpage

  \clearpage
  
  \clearpage

  \clearpage
  
  \clearpage
  
  \clearpage

  \clearpage
  
  \clearpage

  \clearpage

  \clearpage

  \clearpage
  
  \clearpage

  \clearpage

  \clearpage
  
  \clearpage
  
  \clearpage

\end{document}